\documentclass[aps,prb,reprint,onecolumn,11pt,groupedaddress]{revtex4-2}
\usepackage[colorlinks=true,linkcolor=blue,citecolor=blue,urlcolor=blue]{hyperref}
\usepackage{amsmath}
\usepackage{amssymb} 
\usepackage[pdftex]{graphicx}
\usepackage{braket} 

\newcounter{firstbib}

\begin{document}

\newcommand{\RedText}[1]{\textcolor{red}{#1}}
\newcommand{\GreenText}[1]{\textcolor{green}{#1}}
\newcommand{\BlueText}[1]{\textcolor{blue}{#1}}

\title{Storing quantum coherence in a quantum dot nuclear spin ensemble for over 100 milliseconds}


\author{Harry E. Dyte}
\affiliation{Department of Physics and Astronomy, University of
Sheffield, Sheffield S3 7RH, United Kingdom}
\author{Santanu Manna}
\affiliation{Institute of Semiconductor and Solid State Physics,
Johannes Kepler University Linz, Altenberger Str. 69, 4040 Linz,
Austria}
\author{Saimon F. Covre da Silva}
\affiliation{Institute of Semiconductor and Solid State Physics,
Johannes Kepler University Linz, Altenberger Str. 69, 4040 Linz,
Austria}
\author{Armando Rastelli}
\affiliation{Institute of Semiconductor and Solid State Physics,
Johannes Kepler University Linz, Altenberger Str. 69, 4040 Linz,
Austria}
\author{Evgeny A. Chekhovich}
\email[]{E.Chekhovich@sussex.ac.uk} 
\affiliation{Department of
Physics and Astronomy, University of Sussex, Brighton BN1 9QH,
United Kingdom}

\date{\today}

\begin{abstract}
States with long coherence are a crucial requirement for qubits and quantum memories. Nuclear spins in epitaxial quantum dots are a great candidate, offering excellent isolation from external environments and on-demand coupling to optical flying qubits. However, coherence times are limited to $\lesssim1$~ms by the dipole-dipole interactions between the nuclei and their quadrupolar coupling to inhomogeneous crystal strain. Here, we combine strain engineering of the nuclear spin ensemble and tailored dynamical decoupling sequences to achieve nuclear spin coherence times exceeding 100~ms. Recently, a reversible transfer of quantum information into nuclear spin ensembles has been demonstrated in quantum dots. Our results provide a path to develop this concept into a functioning solid-state quantum memory suitable for quantum repeaters in optical quantum communication networks.
\end{abstract}


\maketitle


\section{Introduction}
Quantum memories are indispensable in large scale quantum networks, which are expected to enable long distance communication of quantum information \citep{Ladd_2010, Lvovsky_2009,Pang_2020,Bussières_2014}. Quantum memories have several key requirements \cite{Lvovsky_2009}, a primary figure of merit is the storage time, which is directly related to quantum repeater communication distance. Millisecond-range storage time allows for improvements over direct transmission through an optical fiber \cite{Tittel_2010,Zhao_2008}. Although entanglement generation rate is currently the main limitation \cite{Yu_2020}, its continuous improvement highlights the need for even longer storage times, exceeding 100~ms, in order to achieve worldwide optical communication.

The storage of a quantum state in a memory is limited by the coherence time $T_{2}$. Several material systems offer long $T_{2}$, ranging from seconds to hours \cite{Neuwirth_2021}, including trapped atomic ensembles \cite{Treutlein_2004,Deutsch_2010}, ions \citep{Langer_2005,Bollinger_1991,Wang_2017,Wang_2021,Zhong_2015,Bruzewicz_2019}, electron spins \cite{Tyryshkin_2003,Tyryshkin_2011} and phosphorus nuclear spins \citep{Morton_2008,Steger_2012,Freer_2017} in silicon, as well as electron and nuclear spins of impurities in diamond \citep{Atatre_2018,Stas_2022}. However, long $T_{2}$ are often negated by poor optical properties required for a long-distance quantum network. There are promising hybrid approaches, such as combination of transmon qubits with solid-state quantum memories \cite{Kubo_2011}, but these often suffer from coupling inefficiencies and bandwidth mismatch \cite{Neuwirth_2021} (see further discussion in Supplementary Note~4).

\begin{figure*}
\includegraphics[width=0.98\textwidth]{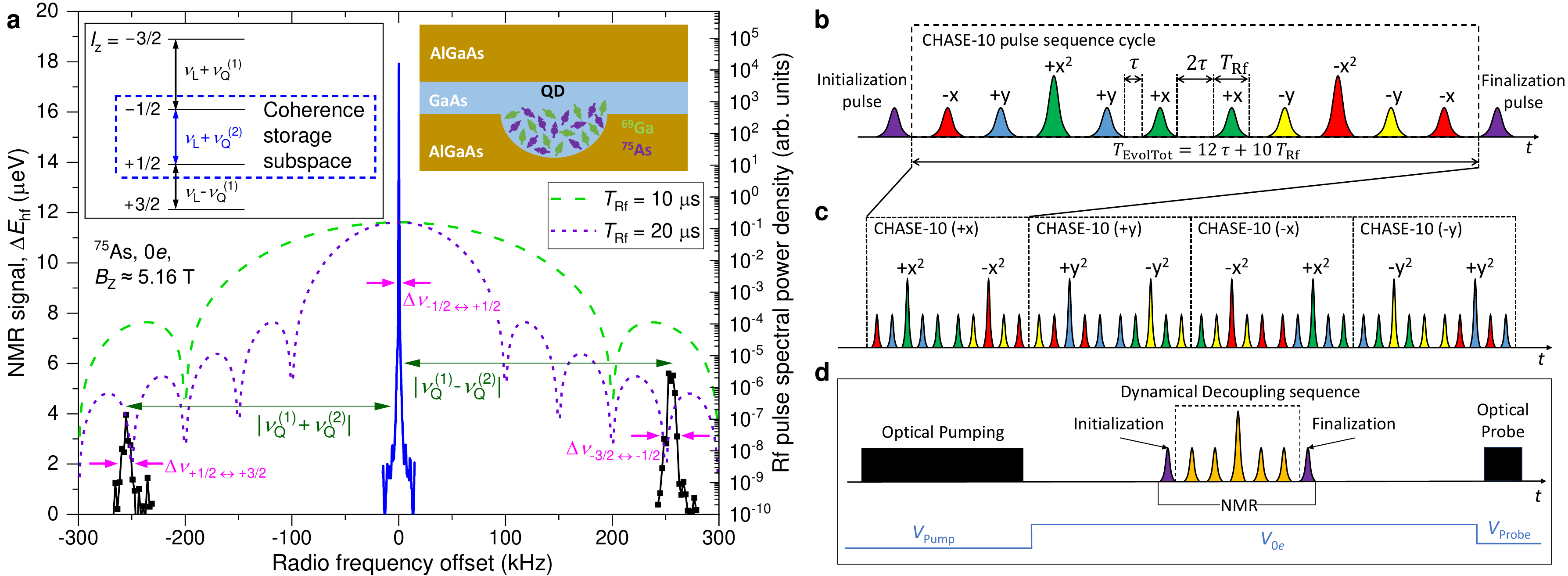}
\caption{\textbf{Optically detected nuclear magnetic resonance of a single quantum dot.} \textbf{a} Right inset shows schematic diagram of $^{69}$Ga and $^{75}$As nuclear spins in a GaAs/AlGaAs QD. NMR spectrum of the spin-3/2 $^{75}$As nuclei (black and blue solid lines, left scale) in an uncharged (0\textit{e}) QD. Frequency offset is shown with respect to the  Larmor frequency $\nu_{L}\approx37.981$~MHz arising from the Zeeman splitting at external field of $B_{z}\approx5.16$~T (left inset). Out of the three magnetic dipole transitions, the two satellite transitions (STs) undergo a first-order quadrupolar shift $\pm\nu_{Q}^{(1)}$ (where $\nu_{Q}^{(1)}\approx255.1$~kHz), while the central transition (CT) is affected only by the second-order quadrupolar shift $\nu_{Q}^{(2)}\approx3.3$~kHz. The CT linewidth ($\Delta\nu_{-1/2\leftrightarrow+1/2}\approx~0.8$~kHz) is much narrower than the ST linewidth ($\Delta\nu_{+1/2\leftrightarrow+3/2}\approx\Delta\nu_{-3/2\leftrightarrow-1/2}\approx13.8$~kHz). Dashed lines (right scale) show spectral profiles of the Rf pulse bursts with duration $T_{\textrm{Rf}}=10$  or 20~$\mu$s, tuned in resonance with the CT. \textbf{b} Schematic diagram of a CHASE-10 sequence cycle, letters and signs denote Rf pulse phases. The pulses are separated by the free-evolution intervals $\tau$. The total nuclear evolution time is $T_{\rm{EvolTot}}=10 T_{\rm{Rf}}+12 \tau$, while $T_{\rm{FreeEvol}}=12\tau$ is the pure free evolution time for one cycle. \textbf{c} The CHASE-40 supercycle constructed of four CHASE-10 steps, with pulse phase incremented by $\pi$/2 in each step. \textbf{d} Timing of the ODNMR measurement cycle. Optical pumping creates longitudinal nuclear spin polarization. The initialization $\pi$/2 Rf pulse converts this into transverse (coherent) nuclear polarization in the $xy$ plane. Dynamical decoupling is applied, followed by a finalization $\pi$/2 pulse to rotate the remaining transverse polarization back along the z-axis. Finally, the nuclear polarization is read out using photoluminescence (PL) spectroscopy under an optical probe pulse. The sample bias is pulsed to maximize optical nuclear spin pumping and PL intensity during optical probing. \label{Fig:Intro}}
\end{figure*}

Epitaxial quantum dots (QDs) in group III-V semiconductors have high qubit entanglement rates \cite{Stockill_2017} and are excellent on-demand emitters of single \cite{Neuwirth_2021,Huber_2017,Schweickert_2018,Arakawa_2020} and entangled photons \cite{Liu_2019,Rota_2024}. At the same time, QDs host material qubits: Electron spin qubits can be interfaced with optical photon qubits \cite{DeGreve_2012,Coste_2023,Laccotripes_2024}, but the coherence of the electron spin is limited to $\approx100~\mu$s \cite{Zaporski_2023}. The nuclear spins are isolated from external environments, resulting in long lifetimes and coherence times \cite{Gillard_2021,Chekhovich_2020,Gillard_2022}. Recently, two-way quantum state transfer between photon and nuclear spins has been demonstrated in a QD, using an electron spin qubit as a mediator \cite{Gangloff_2019,Appel_2024}. However, since all atoms in group III-V materials have non-zero nuclear spins, the natural nuclear spin coherence is limited to a rather modest $\approx1$~ms range \cite{Chekhovich_2015}. Extending nuclear spin coherence is thus a key task in achieving quantum memories suitable for quantum repeaters \cite{Childress_2006,Sharman_2021}.

Here, we achieve nuclear spin coherence of over $T_2\approx100$~ms, made possible by applying two concepts: Firstly, elastic strain is used to engineer the spin-3/2 nuclei and spectrally isolate the subspace with $I_{\rm{z}}=\pm1/2$ spin projections. Small inhomogeneity of this subspace allows application of thousands of coherent control operations, thus enabling efficient dynamical decoupling. Secondly, a dedicated 40-pulse decoupling sequence is designed to extend nuclear spin ensemble coherence while overcoming the parasitic spin locking effects encountered in previous decoupling experiments \cite{Li_2008,Waeber_2019}. Analysis shows that residual decoherence is dominated by the finite-pulse effects and the effective three-body nuclear spin interactions, which are often overlooked. We predict that even longer $T_2$, on the order of $\approx 1$~s, is well within reach through larger strains and further advances in dynamical decoupling. The macroscopically long coherence times achieved here do not rely on isotope enrichment, which is available only in group IV material spin qubits \cite{Balasubramanian_2009,Steger_2012}. Our demonstration of engineered long coherence unlocks the unrivaled optical properties of group III-V materials for applications in quantum memory devices.

\section{Results}
\begin{figure*}[hbt!]
\centering
\includegraphics[width=0.98\textwidth]{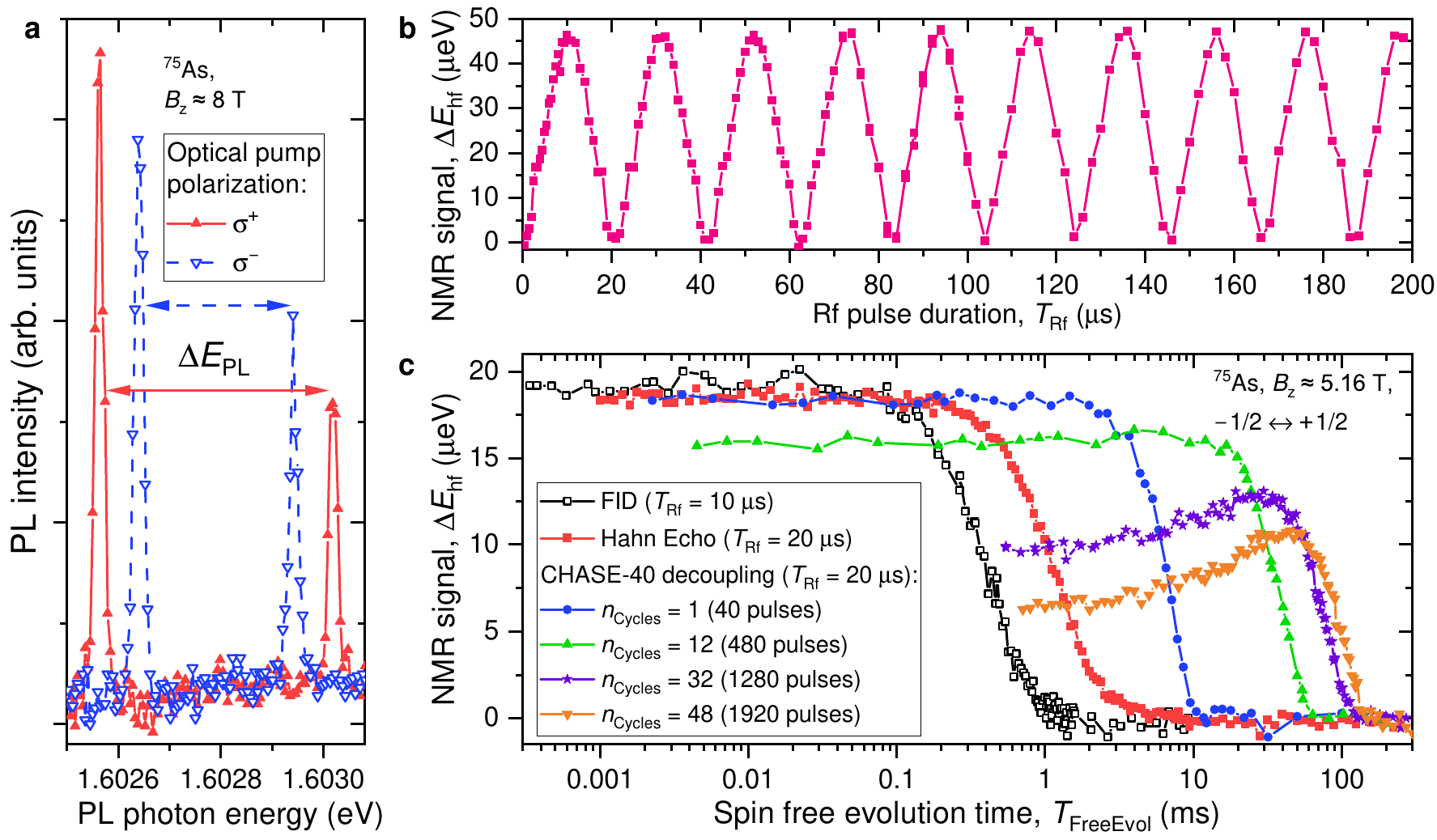}
\caption{\textbf{Dynamical decoupling of QD nuclear spins.} \textbf{a} Two photoluminescence (PL) spectra of a neutral exciton in the same individual QD measured after optical pumping with $\sigma^+$ ($\sigma^-$) polarized light, which results in negative (positive) nuclear spin polarization. PL spectral splitting $\Delta E_{\rm{PL}}$ is a sum of the constant Zeeman splitting and the nuclear hyperfine shift $\Delta E_{\textrm{hf}}$, which is derived from the variations of $\Delta E_{\rm{PL}}$. \textbf{b} Rabi oscillations of the nuclear spins in a neutral (0$e$) QD observed under an increasing duration $T_{\rm{Rf}}$ of an Rf pulse of constant amplitude. \textbf{c} Nuclear spin decoherence measured by optically detected nuclear magnetic resonance (ODNMR) under free induction decay (FID, open squares), Hahn echo (solid squares) and an increasing number of CHASE-40 dynamical decoupling cycles $n_{\textrm{Cycles}}=1 - 48$ (see legend). Rf pulses, with duration $T_{\textrm{Rf}}=20~\mu$s are applied to the central spin transition $-1/2\leftrightarrow+1/2$ in a neutral (0$e$) QD. Nuclear spin polarization is initialized with an x pulse ($\phi=0$).}\label{Fig:ODNMR}
\end{figure*}

\textbf{Strain-engineered nuclear spin ensemble.} We study the nuclear spin coherence of GaAs/AlGaAs QDs grown by molecular beam epitaxy. The right inset of Fig.~\ref{Fig:Intro}a sketches the QD nuclear spin system of $N\approx5\times10^{4}$ nuclei. The three isotopes $^{75}$As, $^{69}$Ga, and $^{71}$Ga all have nuclear spin $I=3/2$. The sample is cooled to $\approx4.2$~K. A superconducting magnet is used to apply a static magnetic field $B_{\rm{z}}\approx 5.16$~T along the sample growth crystal direction [001], lifting the degeneracy of the four nuclear spin states with spin projections $I_{\rm{z}}=\pm1/2,\pm3/2$ (left inset in Fig.~\ref{Fig:Intro}a). Optical pumping with circularly polarized light (Faraday geometry) is used to polarize the nuclear spins along the static magnetic field \cite{Millington-Hotze_2024}. The nuclear spin lifetime is typically $T_{1}>10$~s \cite{MillingtonHotze_2023}, significantly longer than the coherence times measured in this work. Nuclear spin polarization is measured via photoluminescence (PL) spectroscopy \cite{Urbaszek_2013}, see examples in Fig.~\ref{Fig:ODNMR}a. A copper coil generates oscillating magnetic field perpendicular to $B_z$, enabling optically detected nuclear magnetic resonance (ODNMR) measurements. Radio frequency (Rf) bursts with raised cosine envelope are used to transfer coherence in and out of the storage nuclear spin subspace $I_{\rm{z}}=\pm1/2$ and to perform its dynamical decoupling.

The sample is stressed uniaxially along the [110] crystal direction, perpendicular to the static magnetic field. The resulting anharmonicity \cite{Chekhovich_2020,Dyte_2024}, is characterised by the first-order qudrupolar splitting $\nu_{Q}^{(1)}$.  The measured NMR spectrum, shown in Fig.~\ref{Fig:Intro}a for $^{75}$As nuclei in a neutral (0\textit{e}) GaAs/AlGaAs QD, reveals $\nu_{Q}^{(1)}\approx255.1$~kHz. This splitting significantly exceeds the linewdiths of the NMR transitions: the full width at half maximum (FWHM) $\Delta\nu_{+1/2\leftrightarrow+3/2}\approx\Delta\nu_{-3/2\leftrightarrow-1/2}\approx13.8$~kHz of the satellite transitions (STs) $-3/2\leftrightarrow-1/2$ and $+1/2\leftrightarrow+3/2$
is dominated by the inhomogeneous qudrupolar broadening, while the FWHM $\Delta\nu_{-1/2\leftrightarrow+1/2}\approx0.8$~kHz of the central transition $-1/2\leftrightarrow+1/2$ is controlled by a combination of the second-order quadrupolar inhomogeneity and the dipole-dipole interactions \cite{Klauder_1962,Waeber_2019}. The small CT linewidth combined with its strain-induced spectral isolation from STs make it an ideal spin subspace for coherence storage. Notably, the lattice-matched GaAs/AlGaAs QDs offer a significant advantage over Stranski-Krastanov QDs where $\Delta\nu_{-1/2\leftrightarrow+1/2}\approx10 - 40$~kHz \cite{Waeber_2019}.

\textbf{Hamiltonian engineering of a nuclear spin ensemble.} Dynamical control of spin interactions is a powerful technique in magnetic resonance \cite{Haeberlen_1968,Mehring_1983}. The method is based on applying a sequence of Rf pulses that perform fast coherent rotations of the spins, separated by the free evolution intervals. In the interaction picture (the ``toggling'' frame of reference) the Rf pulses can be viewed as transformations of the spin-interaction Hamiltonian. The $\pi$-pulse rotations invert the sign of the frequency shifts, allowing refocusing of the dephasing \cite{Hahn_1950}, which in QDs is caused primarily by inhomogeneous quadrupolar broadening. On the other hand, a sequence of four phase-shifted $\pi$/2-pulses transforms the nuclear spin-spin dipolar interactions in such a way that averages them to zero over the pulse sequence cycle \cite{Waugh_1968}. The average Hamiltonian is the leading (0th order) term in the Magnus expansion of the entire Hamiltonian in the toggling frame. By introducing more complex sequences of pulses it is possible to eliminate the unwanted interactions to higher orders, thus engineering the spin Hamiltonian to have the desired form \cite{Mehring_1983}.

Here, we engineer a “time suspension” \cite{Cory_1990} type of sequence, where the Hamiltonian terms are eliminated as much as possible to preserve an arbitrary coherent
state of the nuclear spin ensemble for
the longest possible time. As a starting point we use a CHASE-10 cyclic sequence of $\pi$/2 and $\pi$ pulses \cite{Waeber_2019} shown in Fig.~\ref{Fig:Intro}b. This sequence eliminates the average (0th order) free-evolution Hamiltonian both for the resonance frequency shifts and the spin-spin interactions. By symmetrizing the sequence, a CHASE-20 supercycle is formed, which further eliminates all the 1st-order terms in the Hamiltonian. The CHASE-20 sequence has been applied to QDs previously, demonstrating its ability to suppress decoherence even under large inhomogeneous resonance broadenings in Stranski-Krastanov InGaAs/GaAs QDs \cite{Waeber_2019}. In low-strain GaAs/AlGaAs QDs, nuclear spin coherence times up to $T_2\approx20$~ms have been achieved \cite{Chekhovich_2020}. However, the $\pi$ pulses cause spin locking \cite{Waeber_2019,Li_2008} which selectively accelerate decoherence of the spin states polarized along a certain equatorial axis of the Bloch sphere in the rotating frame, while artificially enhancing (``locking'') the states polarized along the orthogonal equatorial axis. This behaviour is unwanted in quantum memory applications as it may lead to distortion of the state during storage.

Here we use a different approach, where four CHASE-10 cycles are combined into a CHASE-40 supercycle shown in Fig.~\ref{Fig:Intro}c. The phases of the Rf pulses in each CHASE-10 subcycle are stepped by $\pi$/2. While each subcycle causes spin locking, the preferential direction of the ``lock'', when viewed in the rotating frame, makes a full rotation around the direction of the static magnetic field ($z$) over the CHASE-40 supercycle. This four-step ``rotating spin lock'' eliminates the net spin locking effect for an arbitrary coherent state, as demonstrated through rigorous calculation (See Supplementary Note~5). Furthermore, the leading order residual Hamiltonian of CHASE-40 is twice smaller than in CHASE-20, resulting in extended coherence.

\textbf{Extended spin coherence under dynamical decoupling.} We start by examining experimentally the nuclear spin dynamics under continuous resonant Rf driving. The results shown in Fig.~\ref{Fig:ODNMR}b reveal Rabi oscillations, which confirm the coherent nature of spin driving and allow the $\pi/2$ and $\pi$ Rf pulses to be calibrated for dynamical decoupling (See Supplementary Note~2C). We then apply dynamical decoupling to the isolated $I_{\rm{z}}=\pm1/2$ nuclear spin subspace with two varying parameters: the number of sequence cycles $n_{\rm{Cycles}}$ and the total free evolution time $T_{\rm{FreeEvol}}$. Fig.~\ref{Fig:ODNMR}c shows nuclear spin coherence decay measured using ODNMR. In the simplest case of free induction decay (FID), without any dynamical decoupling (open black squares), the dephasing time is $T_2^*\approx0.46$~ms and is a combined effect of quadrupolar inhomogeneity and dipole-dipole interactions. By using a single $\pi$ pulse as a decoupling sequence (solid squares), we find the Hahn echo \cite{Hahn_1950} coherence time $T_2^{\rm{HE}}\approx1.38$~ms, which is dominated by the dipole-dipole interactions. 

The measured effect of dynamical decoupling with one cycle of CHASE-40 is shown by the circles in Fig.~\ref{Fig:ODNMR}c. The decay of the transverse nuclear polarization is plotted as a function of the total free evolution time $T_{\rm{FreeEvol}}$ (i.e. excluding the duration of the Rf pulses), and reveals a significant extension of the coherence time $T_{2}^{\rm{1xCHASE-40}} \approx12.3$~ms. With the increasing number of CHASE-40 cycles the coherence time is extended further, reaching $T_{2}^{\rm{48xCHASE-40}}\approx106.6$~ms for $n_{\rm{Cycles}}=48$ (orange triangles), an improvement by 2 orders of magnitude compared to the bare Hahn echo coherence time. The achieved nuclear spin coherence is also 3 orders of magnitude longer than the coherence of a dynamically decoupled electron spin in these QDs \cite{Zaporski_2023}. These results demonstrate the superior properties of the nuclear spins as a coherence-storage medium. 

\begin{figure*}
\includegraphics[width=0.99\linewidth]{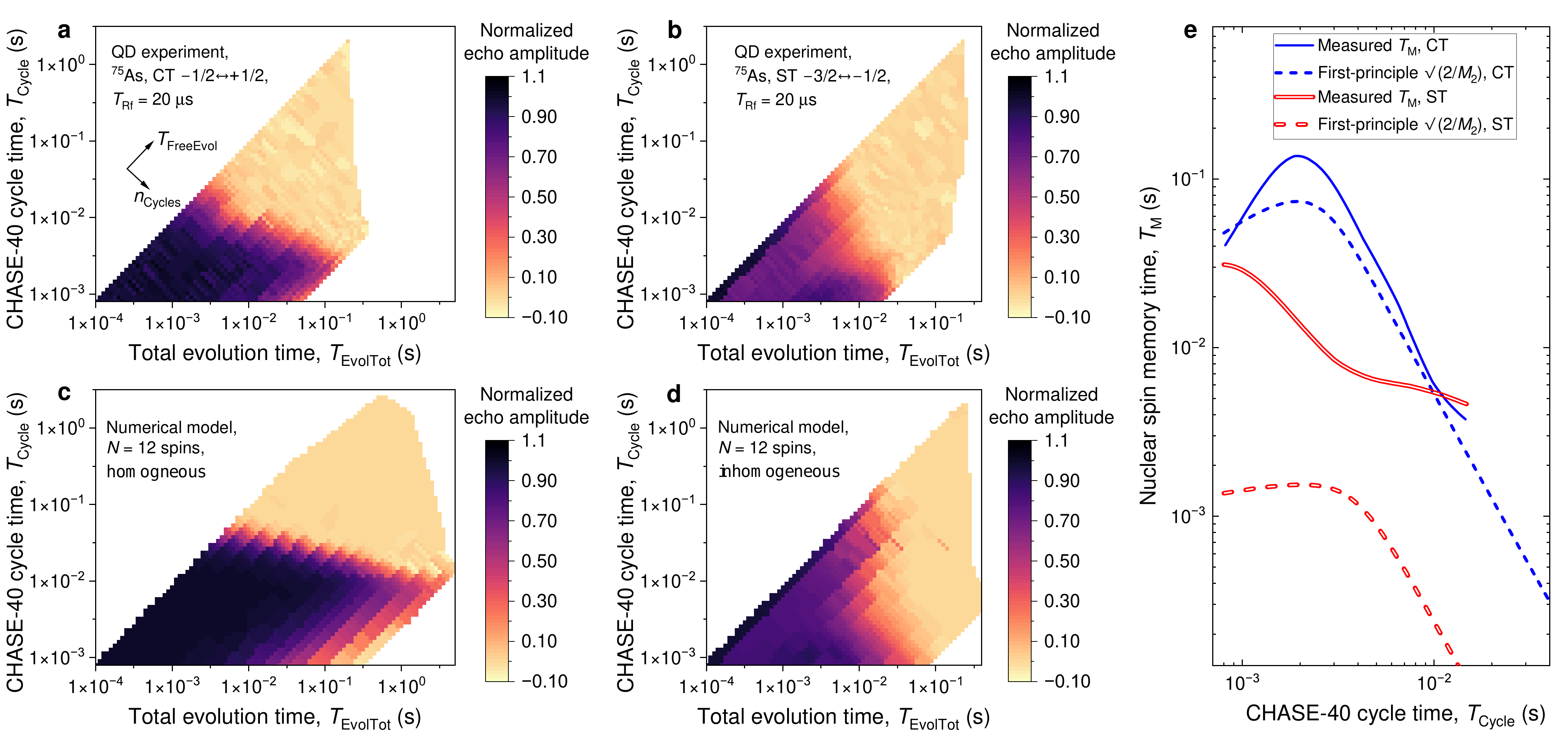}
\caption{\label{Fig:TMem} {\textbf{Coherence storage in a nuclear spin ensemble under dynamical decoupling.}} {\textbf{a}} Decoherence of the $I_{\rm{z}}=\pm1/2$ subspace of the $^{75}$As nuclei measured in a neutral (0$e$) QD under dynamical decoupling with Rf pulse duration of $T_{\rm{Rf}}=20~\mu$s. normalized echo amplitude (colour scale) is shown as a function of the total evolution time $T_{\rm{EvolTot}}=n_{\rm{Cycles}}T_{\rm{Cycle}}$ (free evolution plus Rf pulses, horizontal axis) and the rate of Rf pulsing expressed in terms of the duration $T_{\rm{Cycle}}=40T_{\rm{Rf}}+48\tau$ of a single CHASE-40 cycle (vertical axis). The axes of the plot correspond to a nonlinear transformation from the variables $n_{\rm{Cycles}}$ and $T_{\rm{FreeEvol}}=40 n_{\rm{Cycles}}\tau$ used in Fig.~\ref{Fig:ODNMR}c. The plot combines data obtained with different numbers of CHASE-40 cycles and with shorter subcycles such as CHASE-10. {\textbf{b}} Decoherence of the $I_{\rm{z}}=(-3/2,-1/2)$ subspace of the $^{75}$As nuclei measured under dynamical decoupling with $T_{\rm{Rf}}=20~\mu$s. {\textbf{c}} Decoherence under dynamical decoupling derived from first-principles numerical modeling of spin dynamics of a homogeneous ensemble of $N=12$ nuclei. {\textbf{d}} Numerically modeled decoherence of $N=12$ nuclei subject to inhomogeneous spectral broadening. {\textbf{e}} Nuclear spin memory time $T_{\rm{M}}$ derived from exponential fitting of decoherence measured as a function of $T_{\rm{EvolTot}}$. Single and double solid lines show results for $I_{\rm{z}}=\pm1/2$ CT and $I_{\rm{z}}=(-3/2,-1/2)$ ST subspaces of $^{75}$As nuclei, respectively. $T_{\rm{M}}$ calculated analytically from the second moment $M_2$ of the residual Hamiltonian under dynamical decoupling are shown by the single and double dashed lines for $I_{\rm{z}}=\pm1/2$ and $I_{\rm{z}}=(-3/2,-1/2)$ subspaces, respectively.}
\end{figure*}

It can be seen from Fig.~\ref{Fig:ODNMR}c that under an increasing number
of decoupling cycles $n_{\rm{Cycles}}$ the coherence (the NMR echo signal) is reduced even in the limit of $T_{\rm{FreeEvol}}\rightarrow0$. Moreover, the dependence of transverse nuclear spin polarization on $T_{\rm{FreeEvol}}$ becomes nonmonotonic. These are the indications of nuclear spin decoherence during the finite (nonzero duration) Rf pulses. For quantum memory applications, we are interested in minimising decoherence, whether caused by free evolution or the Rf control pulses. We seek this optimum by replotting the data of Fig.~\ref{Fig:ODNMR}c in Fig.~\ref{Fig:TMem}a, where the normalized nuclear spin coherence is shown as a function of the total evolution (free evolution plus Rf pulses) time $T_{\rm{EvolTot}}$ (horizontal axis) and the duration $T_{\rm{Cycle}}$ of one CHASE-40 cycle (vertical axis). The individual decay plots of Fig.~\ref{Fig:ODNMR}c measured at fixed $n_{\rm{Cycles}}$ now appear along the diagonal lines in Fig.~\ref{Fig:TMem}a. We fit the decay of coherence with an exponential function of $T_{\rm{EvolTot}}$: the resulting decay time, which we denote as spin memory time $T_{\rm{M}}$, is distinct from coherence time $T_{2}$ and is shown as a function of $T_{\rm{Cycle}}$ by the single solid line in Fig.~\ref{Fig:TMem}e. The maximum $T_{\rm{M}}\approx136$~ms is achieved not under the fastest possible Rf pulsing, but at $T_{\rm{Cycle}}\approx2$~ms, which is 5 times longer than the minimum $T_{\rm{Cycle}}\approx0.8$~ms achieved at $T_{\rm{Rf}}=20~\mu$s. This confirms that the finite-pulse effects are the main limitation to extending coherence storage through fast dynamical decoupling.

Decoherence during the Rf pulses can in principle be suppressed by reducing $T_{\rm{Rf}}$. However, experiments conducted on the $I_{\rm{z}}=\pm1/2$ spin states with a reduced $T_{\rm{Rf}}=10~\mu$s yield faster decoherence than under $T_{\rm{Rf}}=20~\mu$s (see Supplementary Note~3A). This seemingly contradictory result is understood by considering the spectral profiles of the Rf pulses (dashed lines in Fig.~\ref{Fig:Intro}a). While the unwanted spin decoherence is indeed suppressed under the shorter $T_{\rm{Rf}}=10~\mu$s pulses, their broader spectral profile results in a stronger overlap with the STs. Such overlap leads to a faster ``leakage'' of coherence from the storage $I_{\rm{z}}=\pm1/2$ subspace into the $I_{\rm{z}}=\pm3/2$ states. Thus, there is an optimal pulse duration that balances the finite-pulse and the leakage effects. For the studied structure with the strain-induced quadrupolar splitting of $\nu_{\rm{Q}}^{(1)}\approx250$~kHz, this optimum is close to $T_{\rm{Rf}}=20~\mu$s. Increasing elastic strain from $\approx0.0025$ in the studied sample to the $\approx0.01$ range \cite{Huo_2013} is a promising route for applying shorter Rf pulses and a further significant improvement of the storage time and fidelity in a QD nuclear spin quantum memory.

\textbf{CHASE decoupling of an inhomogeneous spin ensemble.} In order to demonstrate the importance of isolating the homogeneous $I_{\rm{z}}=\pm1/2$ CT storage subspace, we examine the opposite case of a $I_{\rm{z}}=(-3/2,-1/2)$ subspace. The considerably larger inhomogeneous broadening is characterised by the spectral shape of the ST NMR transitions. This is a weighted sum of a visible peak (Fig.~\ref{Fig:Intro}a) with a linewidth of $\Delta\nu_{+1/2\leftrightarrow+3/2}\approx\Delta\nu_{-3/2\leftrightarrow-1/2}\approx~13.8$~kHz (65\% weight), and a much broader invisible peak, which was shown previously \cite{Zaporski_2023} to stretch to $\approx\pm100$~kHz (35\% weight) and is caused by the atomic-scale strain of the randomly positioned Al and Ga atoms. The measured ST decoherence under CHASE decoupling is shown in Fig.~\ref{Fig:TMem}b, and the resulting spin memory time $T_{\rm{M}}$ is shown by the double solid line in Fig.~\ref{Fig:TMem}e. Unlike with $I_{\rm{z}}=\pm1/2$, the best possible dynamical decoupling of the $I_{\rm{z}}=(-3/2,-1/2)$ subspace is achieved at the shortest possible $T_{\rm{Cycle}}$. Despite this fastest possible Rf pulsing, the maximum achieved memory time is $T_{\rm{M}}\approx31$~ms. Although this storage time is a factor of $\approx15$ improvement over the simple Hahn echo, it is a factor of $\approx4.4$ worse than $T_{\rm{M}}$ achieved for the spectrally narrow $I_{\rm{z}}=\pm1/2$ subspace. 

The inferior spin memory time for an inhomogeneously broadened spin ensemble demonstrates how CHASE-40, as any other dynamical decoupling protocol, reaches the limit of its performance when the interaction that is being decoupled is no longer a small perturbation. The inhomogeneous broadening of the $I_{\rm{z}}=\pm1/2$ CT subspace is a small perturbation, characterized by $\Delta\nu_{-1/2\leftrightarrow+1/2}T_{\rm{Rf}}\approx0.015\ll1$. By contrast, the relative broadening of the $I_{\rm{z}}=(-3/2,-1/2)$ subspace is comparable to unity for the visible part of the ST NMR peak ($\Delta\nu_{-3/2\leftrightarrow-1/2}T_{\rm{Rf}}\approx0.28$) and violates the perturbative approximation ($\Delta\nu_{-3/2\leftrightarrow-1/2}T_{\rm{Rf}}>1$) for the broad component of the ST. The large inhomogeneity of the ST exacerbates decoherence through spin evolution during the finite Rf pulses. Moreover, $\Delta\nu_{-3/2\leftrightarrow-1/2}T_{\rm{Rf}}\gtrsim 1$ means that Rf control pulses become more ``soft'' (i.e. not infinitely broad spectrally), resulting in imperfect rotations of the nuclear spins \cite{Khodjasteh_2005,Souza_2012}. On the other hand, CHASE-40 shows no sign of spin-locking even for the $I_{\rm{z}}=(-3/2,-1/2)$ subspace, despite its larger inhomogeneous broadening (See Supplementary Note~3B).

\begin{figure*}[hbt!]
\centering
\includegraphics[width=0.98\textwidth]{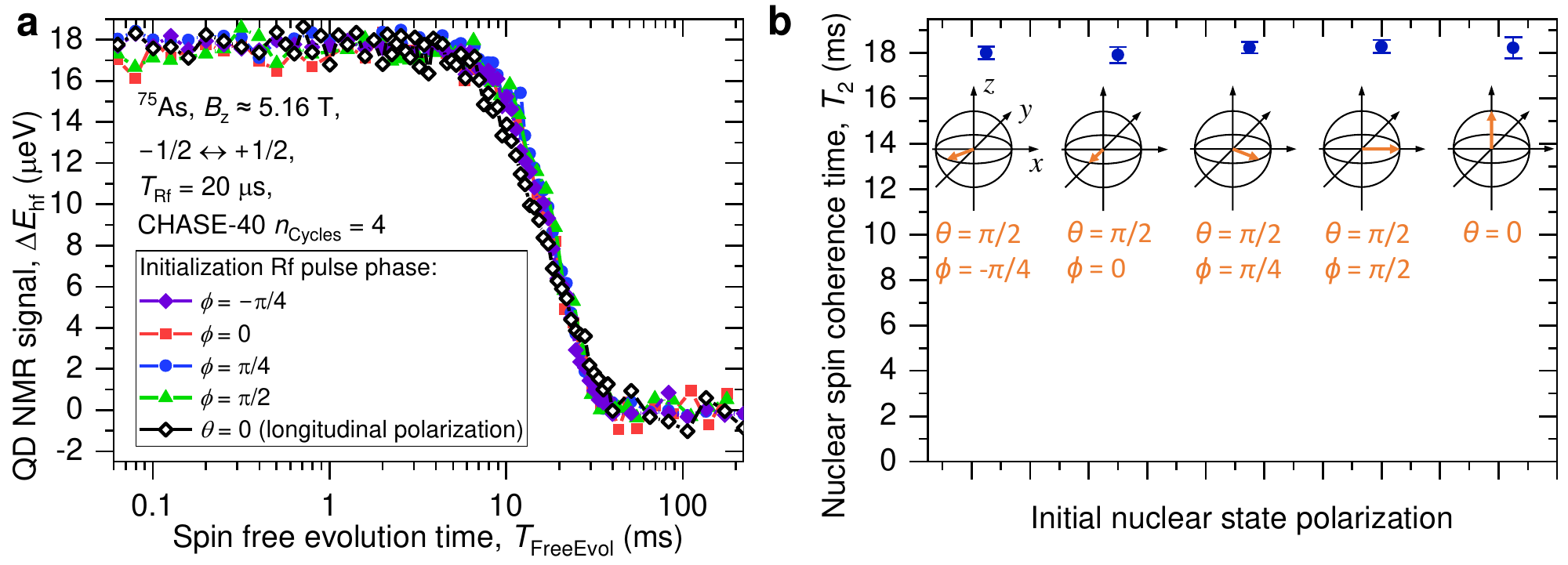}
\caption{\textbf{Uniform decoupling of an arbitrary coherent nuclear spin state.} \textbf{a} Nuclear spin decoherence measured under 4 cycles of CHASE-40 with different phases $\phi$ of the initialization Rf pulse, which initializes transverse nuclear polarization along different azimuth angles in the $xy$ plane ($\theta=\pi/2$). The measurement of the longitudinal nuclear spin relaxation under CHASE-40 (without the initialization pulse, $\theta=0$) is shown by the open diamonds. \textbf{b} The nuclear spin decay times $T_{2}$ and $T_{1}$ obtained from fitting the data in \textbf{a}. Schematics show orientation of the nuclei on the Bloch sphere after the initialization pulse, where present. \label{Fig:PhaseDep} }
\end{figure*}

\textbf{Uniform decoupling of an arbitrary coherent nuclear spin state.} An ideal quantum memory must store any given state with equally high fidelity. However, dynamical decoupling can create parasitic spin locking regimes, where storage effectiveness depends on the initial state \cite{Li_2008,Waeber_2019}. We examine the uniformness of the quantum state storage by measuring dynamical decoupling of the homogeneous $I_{\rm{z}}=\pm1/2$ subspace under four cycles of CHASE-40 with different initial states. The phase $\phi$ of the initialization Rf pulse is varied to prepare transverse nuclear spin polarization along the different axes in the equatorial $xy$ plane of the rotating frame ($\phi=0$ corresponds to an ``+x'' Rf pulse and prepares polarization along the $-y$ axis, a ``+y'' pulse with $\phi=\pi/2$ prepares polarization along the $x$ axis). We further perform a measurement, where the initialization pulse is omitted, corresponding to initial nuclear spin polarization along the strong magnetic field ($\theta=0$). The measured decay curves are shown in Fig.~\ref{Fig:PhaseDep}a. The coherence times obtained from fitting are shown in Fig~\ref{Fig:PhaseDep}b, and are around $T_{2}\approx18$~ms, nearly independent of the initial state.


The uniform dynamical decoupling of different initial states confirms experimentally the design principle of the CHASE-40 supercycle. Cycling of the Rf pulse phases in the four constituent CHASE-10 subcycles can be understood intuitively as a ``rotary spin lock'': the axis of spin locking is slowly precessing with respect to the rotating frame, with a net effect of removing any preferential spin locking axis over the entire CHASE-40 cycle. Rigorous calculations confirm this result, showing that the symmetry axis of the residual Hamiltonian of CHASE-40 is along the strong static magnetic field ($z$ axis). It is also worth noting that the $T_2$ decay time of the transverse polarization (measured with an initialization $\pi/2$ pulse) is nearly identical to that of the longitudinal $T_1$ decay time (measured without any initialization pulse): a well-designed time-suspension sequence eliminates all the leading interaction terms, effectively removing the distinction between the longitudinal ($T_1$) and transverse ($T_2$) relaxation timescales.

The rotary spin lock offers a simple and reliable approach for reusing the dynamical decoupling sequences where spin locking is otherwise present \cite{Dementyev_2003,Li_2007,Waeber_2019}. The robustness of CHASE-40 against spin locking is key to achieving long spin memory times $T_{\rm{M}}\gtrsim100$~ms through repeated cycling (up to 2400 pulses).

\textbf{Predicting dynamical decoupling performance through analytical and numerical modeling.} The effective spin Hamiltonian under dynamical decoupling can be calculated as a Magnus expansion series. However, finding the spin dynamics from a known Hamiltonian is still a difficult problem. On the other hand, the exact spin dynamics is related to the exact NMR spectral lineshape through Fourier transform. The NMR lineshape can be approximated as a Gaussian, and its linewidth can be approximated in terms of the second moment $M_2$, which in turn can be found from the Hamiltonian through direct calculation. The approximate coherent memory time can then be found as $T_{\rm{M}}\approx\sqrt{2/M_2}$ \cite{Mehring_1983}. The residual Hamiltonian of CHASE-40 has been computed analytically up to second order: it is too bulky to reproduce in full, a more detailed discussion can be found in Supplementary Note~5. The Hamiltonian depends on two parameters: the magnitude of the dipole-dipole interaction and the quadrupolar inhomogeneity. These are derived from FID and Hahn Echo experimental data, allowing $T_{\rm{M}}^{\rm{CHASE-40}}$ to be calculated up to second order in analytical form and without any fitting parameters. The results are shown by the dashed lines in Fig.~\ref{Fig:TMem}e. For the homogeneous subspace $I_{\rm{z}}=\pm1/2$ (single dashed line), the analytical model accurately predicts the peak in the spin memory time $T_{\rm{M}}$ at $T_{\rm{Cycle}}\approx2$~ms. The actual peak value of $T_{\rm{M}}$ is underestimated but matches the experiment within a factor of $\approx2$. By contrast, for the inhomogeneous subspace $I_{\rm{z}}=(-3/2,-1/2)$, the model underestimates $T_{\rm{M}}$ by an order of magnitude. This indicates the limited accuracy of the perturbative Magnus expansion, which breaks down when the inhomogeneous broadening (in frequency units) is no longer small when compared to the reciprocal cycle time $1/T_{\rm{Cycle}}$. In principle, the analytical model could be improved by extending the Magnus expansion. However, the increasing complexity of the high-order terms makes this approach impractical.

The advantage of the analytical model is that it allows insights into the underlying phenomena. In particular, we find that direct dipole-dipole interactions of the $i$th and $j$th spins of the ensemble, characterised by the coupling constant $\nu_{ij}$, is eliminated by CHASE-40 up to second order inclusive. However, the dipolar interaction of the spins $i, j$ remains, but with a coupling constant $\propto\nu_{ik}\nu_{kj}$, where $k\ne i,j$ is any other spin. Such a term can be interpreted as a three-particle coupling, where the interaction of any two spins $i$ and $j$ is mediated by any other spin $k$. Although often ignored, here we find that the residual decoherence of the homogeneous subspace  $I_{\rm{z}}=\pm1/2$ can be explained only by taking into account this effective three-body interaction. The three-body second-order interaction limits $T_{\rm{M}}$ under slow dynamical decoupling (long $T_{\rm{Cycle}}$). In the oposite limit of fast decoupling (short $T_{\rm{Cycle}}$) the memory time $T_{\rm{M}}$ is limited by the zero-order dipole-dipole term arising from finite $T_{\rm{Rf}}>0$. A combination of these two effects results in a non-monotonic dependence $T_{\rm{M}}(T_{\rm{Cycle}})$ with a maximum in $T_{\rm{M}}$, as shown by the single lines in Fig.~\ref{Fig:TMem}e. By contrast, the decoherence in the inhomogeneous subspace $I_{\rm{z}}=(-3/2,-1/2)$ is dominated by the quadrupolar offset inhomogeneity under finite ($T_{\rm{Rf}}>0$) control pulses.

We further conduct numerical modeling of the CHASE dynamical decoupling by solving the exact Schrodinger equation of a system of $N=12$ spins (see details in Supplementary Note 6). The results for the case of the inhomogeneous subspace $I_{\rm{z}}=(-3/2,-1/2)$, are shown in Fig.~\ref{Fig:TMem}d. Since the decoherence rate is dominated by the quadrupolar offsets, which is a single-particle effect, a good quantitative agreement with the experiment (Fig.~\ref{Fig:TMem}b) is obtained by using realistic values of the inhomogeneous quadrupolar shifts in the numerical model. The results for the $I_{\rm{z}}=\pm1/2$ subspace, where quadrupolar inhomogeneity is taken to be zero, are shown in Fig.~\ref{Fig:TMem}c. The numerical model reproduces the main features of the experimental data on the $I_{\rm{z}}=\pm1/2$ subspace  (Fig.~\ref{Fig:TMem}a), in particular the nonmonotonic dependence of the spin memory time $T_{\rm{M}}$ on the decoupling sequence cycle time $T_{\rm{Cycle}}$. However, the agreement is only within an order of magnitude: the numerically-simulated maximum $T_{\rm{M}}\approx2$~s occurs at $T_{\rm{Cycle}}\approx8$~ms, compared to the measured maximum $T_{\rm{M}}\approx0.136$~s at $T_{\rm{Cycle}}\approx2$~ms. This discrepancy may seem unexpected, given that the numerically-simulated coherence time under simple Hahn echo $T_2^{\rm{HE}}\approx1.4~$ms is very close to the measured $T_2^{\rm{HE}}\approx1.38~$ms. However, the decoherence under Hahn echo is governed by the direct (pairwise) dipole-dipole interaction of the nuclear spins, whereas decoherence under CHASE-40 is dominated by the effective three-spin interaction. We therefore ascribe the discrepancy in $T_{\rm{M}}$ to the limited number of spins in the numerical model: the number of three-spin combinations contributing to decoherence in an ensemble with $N=12$ is considerably smaller than in a real crystal lattice of a QD. The discrepancy is also likely to include the small but nonzero quadrupolar inhomogeneity of the $I_{\rm{z}}=\pm1/2$ subspace, which reduces CHASE-40 $T_{\rm{M}}$ in a real QD.    

\section{Discussion}

We have demonstrated very long coherence storage times of over 100~ms in a nuclear spin ensemble of an optically-active GaAs semiconductor QD. These QDs are a promising candidate for quantum memory, which integrates the spin qubit with a single photon sources \cite{Neuwirth_2021}, thus avoiding the need for complex hybrid schemes \cite{Kubo_2011}. Long-term preservation of nuclear spin coherence demonstrated here is a key step in bringing the concept of QD-based optical quantum memory \cite{Appel_2024} to practical implementation. The extended coherence is enabled by strain engineering of the nuclear spin ensemble and the tailored 40-pulse time-suspension decoupling sequences.

We use a three-pronged approach to the design of dynamical decoupling sequences. Numerical modelling can predict the overall performance of a dynamical decoupling protocol, but its accuracy is limited by the small number of spins $N$, constrained in turn by the exponential scaling of the required computing resources with increasing $N$. As a result, numerical simulations are time-consuming: full datasets of Figs.~\ref{Fig:TMem}c, d require many days of computations on a workstation PC, which is comparable to experimental time required for Figs.~\ref{Fig:TMem}a, b. Analytical calculations provide good predictions in case of small inhomogeneity. However, for a sequences with 40 pulses, derivation of the residual Hamiltonian takes several hours of computer-assisted algebraic derivations and further tedious manual work to analyse the bulky analytical results. Thus, the two modeling approaches encounter their different limitations, leaving experiment as the ultimate verification of the excellent coherence protection achieved with CHASE-40. The sequence is robust against spin locking, errors in control pulses, and is applicable to spin ensembles with a substantial inhomogeneous broadening.

Strain engineering is a key enabling technique, as it allows spectral isolation of the homogeneous $I_{\rm{z}}=\pm1/2$ nuclear spin subspace. A further increase of elastic strain by a factor of $\approx4$ is within the yield strain of GaAs and is feasible using membrane microstructures \cite{Hjort1994, Huo_2013, MartinSanchez2017}. This would allow quadrupolar splitting in excess of $\nu_{Q}^{(1)}\gtrsim 1$~MHz, enabling further improvement in quantum memory storage time and fidelity. More importantly, the MHz-range quadrupolar splitting would be sufficient to exceed the electron-nuclear hyperfine interaction, which is $\lesssim 200$~kHz in the studied GaAs QDs \cite{Dyte_2024}. While dynamical decoupling of nuclear spins in presence of the central electron spin qubit is possible in principle \cite{Gillard_2022}, achieving long nuclear spin coherence in presence of electron would require spectral isolation (through increased strain) of the hyperfine-broadened nuclear spin transitions. 

The maximum storage time $T_{\rm{M}}$ under CHASE-40 decoupling is limited by the finite duration of the control pulses, which causes a drop in $T_{\rm{M}}$ in the limit of frequent Rf pulsing (single lines in Fig.~\ref{Fig:TMem}e, limit of small $T_{\rm{Cycle}}$). By eliminating the effect of finite pulses \cite{Haeberlen_1968, Burum1979} it should be possible to achieve $T_{\rm{M}}\approx1$~s even at the current level of elastic strain, limited only by the second-order three-particle spin-spin interactions. More broadly, dynamical decoupling can be used to study spin-spin entanglement, thermalization in disordered quantum systems, and many-body localization \cite{Wei_2018,Lukin_2019}.

\begin{acknowledgments}
{\it Acknowledgements:} H.E.D. was supported by an EPSRC doctoral training grant. E.A.C. was supported by a Royal Society University Research Fellowship, the Leverhulme Trust grant RPG-2023-141, and the EPSRC award EP/V048333/1. A.R. acknowledges support of the Austrian Science Fund (FWF) via the Research Group FG5, I 4320, I 4380, I 3762, the Linz Institute of Technology (LIT), and the LIT Secure and Correct Systems Lab, supported by the State of Upper Austria, the European Union's Horizon 2020 research and innovation program under Grant Agreements No. 899814 (Qurope), No. 871130 (Ascent+), the QuantERA II project QD-E-QKD and the FFG (grant No. 891366). A.R. and E.A.C. were supported by the QuantERA award MEEDGARD. {\it Author contributions:} S.M., S.F.C.S. and A.R. developed, grew and processed the quantum dot samples. H.E.D, conducted the experiments. H.E.D. and E.A.C. analysed the data. H.E.D. and E.A.C. drafted the manuscript with input from all authors. E.A.C. performed numerical modelling and coordinated the project.
\end{acknowledgments}


\pagebreak


\renewcommand{\thesection}{Supplementary Note \arabic{section}}
\setcounter{section}{0}
\renewcommand{\thefigure}{\arabic{figure}}
\renewcommand{\figurename}{Supplementary Figure}
\setcounter{figure}{0}
\renewcommand{\theequation}{S\arabic{equation}}
\setcounter{equation}{0}
\renewcommand{\thetable}{\arabic{table}}
\renewcommand{\tablename}{Supplementary Table}
\setcounter{table}{0}

\makeatletter
\def\l@subsubsection#1#2{}
\makeatother

\pagebreak \pagenumbering{arabic}
\newpage


\section*{Supplementary Information}

\section{Sample structure}

\begin{figure}
    \includegraphics[]{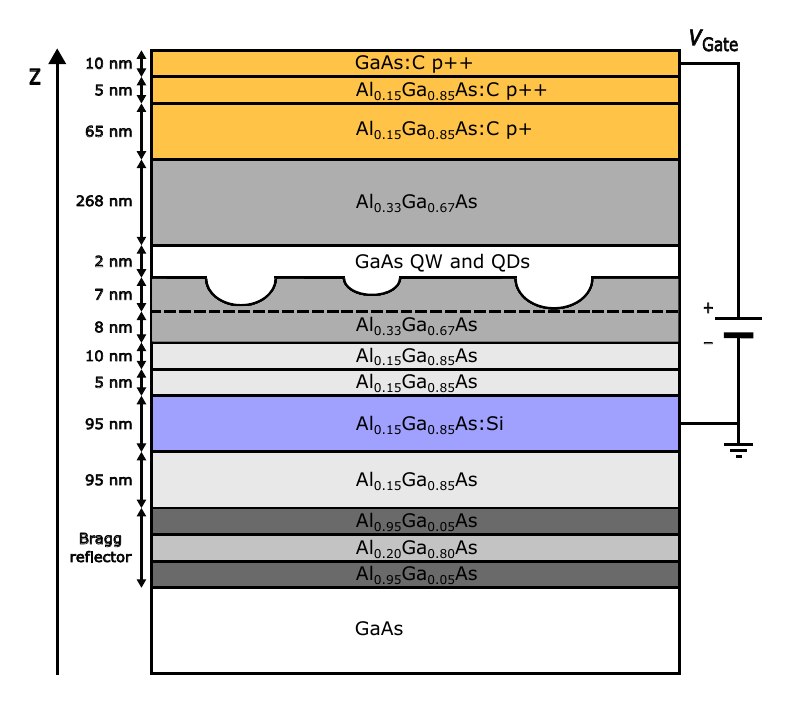}
\caption{Schematic of the quantum dot sample structure.}
\label{SupFig1:Sample}
\end{figure}

The sample used for this work is the same as used in Refs. \cite{MillingtonHotze_2023,Zaporski_2023,MillingtonHotze_2024,Dyte_2024}. The sample is grown on a semi-insulating GaAs (001) substrate. Supplementary Fig.~\ref{SupFig1:Sample} shows the layer sequence of the semiconductor structure. Growth starts with a layer of Al$_{0.95}$Ga$_{0.05}$As followed by a single pair of Al$_{0.2}$Ga$_{0.8}$As and Al$_{0.95}$Ga$_{0.05}$As layers, which act as a Bragg reflector in optical experiments. A 95~nm thick layer of Al$_{0.15}$Ga$_{0.85}$As is then grown followed by a 95~nm thick layer of Al$_{0.15}$Ga$_{0.85}$As doped with Si at a volume concentration of $1.0~\times~10^{18}$~cm$^{-3}$. The concentration of Al is kept at a low value of 0.15 in the Si doped layer, in order to avoid the formation of the deep DX centers \cite{Oshiyama1986,Mooney1990,Zhai2020}. The \textit{n}-type doped layer is followed by the electron tunnel barrier layers: first a 5~nm thick Al$_{0.15}$Ga$_{0.85}$As layer is grown at a reduced temperature of 560~$^{\circ}$C to suppress Si segregation, followed by a 10~nm thick Al$_{0.15}$Ga$_{0.85}$As and then a 15~nm thick Al$_{0.33}$Ga$_{0.67}$As layer grown at 600~$^{\circ}$C. Droplets of Aluminium are grown on the surface of the Al$_{0.33}$Ga$_{0.67}$As layer and are used to etch the nanoholes \cite{Heyn2009,Atkinson2012,Huo_2013APL}. Atomic force microscopy shows typical nanoholes have a depth of $\approx$~6.5~nm and are $\approx$~70~nm in diameter \cite{MillingtonHotze_2023}. A 2.1~nm thick layer of GaAs is grown to form QDs by infilling the nanoholes and additionally form the quantum well (QW) layer. Thus, the maximum height of the QDs in the growth \textit{z} direction is $\approx$~9~nm. The GaAs layer is followed by a 268~nm thick Al$_{0.33}$Ga$_{0.67}$As barrier layer. Finally, the \textit{p}-type contact layers doped with C are grown: a 65~nm thick layer of Al$_{0.15}$Ga$_{0.85}$As with a $5.0~\times~10^{18}$~cm$^{-3}$ doping concentration, a 5~nm thick layer of Al$_{0.15}$Ga$_{0.85}$As with a $9.0\times10^{18}$~cm$^{-3}$ concentration, and a final 10~nm thick layer of GaAs with a $5.0\times10^{18}$~cm$^{-3}$ concentration.

Next, the sample is processed into a \textit{p-i-n} diode structure. Mesa structures with a height of 250~nm are created by etching away the \textit{p}-doped layers and depositing onto the etched regions the following sequence of layers: Ni(10~nm), AuGe(150~nm), Ni(40~nm), Au(100~nm). The sample is then annealed to enable diffusion of the deposited metals down to the \textit{n}-doped layer to form the ohmic back contact. Depositing Ti(15~nm)/Au(100~nm) on to the \textit{p}-type surface of the mesa areas forms the top gate contact. QD photoluminescence (PL) is excited and collected through the top of the sample. Sample gate bias $V_{\rm{Gate}}$ is the bias of the \textit{p}-type top contact with respect to the grounded \textit{n}-type back contact. Tunneling of holes is suppressed due to the large thickness of the top Al$_{0.33}$Ga$_{0.67}$As layer, whereas tunnel coupling to the \textit{n}-type layer enables deterministic charging of the QDs with electrons by changing $V_{\rm{Gate}}$. For this work however, we leave the QD uncharged by applying reverse bias.

To allow the quadrupolar components of the nuclear magnetic resonance (NMR) spectra to be resolved, the semiconductor sample used in this work is subjected to uniaxial mechanical stress. To this end, the semiconductor wafer is first cleaved into a small piece with a rectangular surface area of 0.7~mm$\times$2.35~mm. The edges of the rectangular profile are aligned along the [110] and [1$\bar{1}$0] crystallographic directions. The sample thickness along the [001] growth direction is 0.35~nm. Thus, the sample is a parallelepiped. The sample is then inserted into a home-made stress cell. This is done in such a way that the two 0.7~mm$\times$0.35~mm surfaces of the sample  are contacted to the flat titanium surfaces of the stress cell bracket. Finally, a titanium screw is directed along the 2.35~mm long edge of the sample in order to apply compressive stress \cite{Dyte_2024}.

\section{Experimental techniques}

The sample is placed in a bath-cryostat and cooled using liquid He to $\approx4.2$~K. An inbuilt superconducting coil is used to apply a static magnetic field $B_{z}$ up to 8~T along the \textit{z}-axis parallel to the sample growth direction [001] and the optical axis (Faraday geometry). Therefore, the applied mechanical stress is perpendicular to the field and optical axis. Optical measurements are conducted using a confocal microscopy configuration. An aspheric lens with a focal distance of 1.45~mm and NA~=~0.58 is used as an objective for optical excitation of the QD and for collection of photoluminescence (PL). The excitation laser is focused into a spot with a diameter of $\approx1~\mu$m. A two-stage Czerny-Turner spectrometer is used to analyze the collected PL. The light is collimated and directed to a plane diffraction grating at each stage. Upon dispersion, the light is focused by a mirror with a 1~m focal length. After the spectrometer, a pair of achromatic lens doublets transfer the spectral image onto a charge-coupled device (CCD) photo-detector with a magnification of 3.75. The changes in the spectral splitting, determined from the PL spectra of a neutral exciton $X^{0}$, allow measurement of the hyperfine shifts $E_{\rm{hf}}$, which is proportional to the nuclear spin polarization degree (Fig.~2a of the main text). Both the pump and probe laser pulses are formed using mechanical shutters. One more mechanical shutter is used to block the input of the spectrometer during the optical pumping.

\subsection{Optical pumping of nuclear spin polarization}

\begin{figure}
    \centering
    \includegraphics[width=0.99\linewidth]{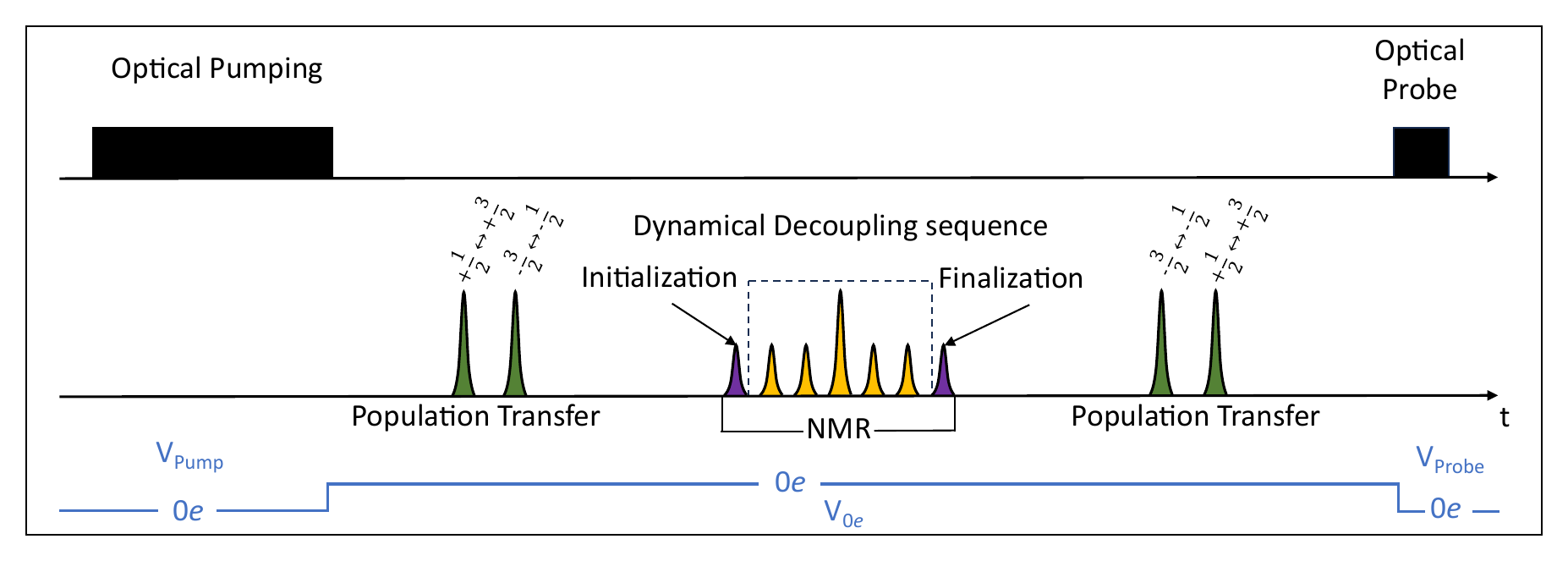}
    \caption{Full timing diagram of pulsed dynamical decoupling experiments.}
    \label{SupFig2:MmtCycle}
\end{figure}

Supplementary Fig.~\ref{SupFig2:MmtCycle} shows the full timing diagram for the NMR measurements used in this work. In the first part of the measurement cycle dynamical nuclear spin polarization is created by optically pumping the QD with a tunable single-mode circularly polarized diode laser. This is a well established technique and allows nuclear spin polarization greater than 50\% to be reached \cite{Ulhaq_2016,Gammon_2001,Eble_2006,SkibaSzymanska_2008,Ragunathan_2019,Urbaszek_2013}. The process is cyclic with three stages. First, optical excitation creates a spin polarized electron, a process enabled by the selection rules in III-V semiconductors allowing spin-polarized electron-hole pairs to be formed from the conversion of circularly polarized light. Secondly, the flip-flop term of the electron-nuclear hyperfine Hamiltonian allows the electron to exchange its spin with a single nuclei. Finally, electron-hole recombination removes the flipped electron allowing another polarized electron to be created in the dot and the process repeats, building-up polarization of the ensemble of nuclei. The pump laser pulse is typically 5~s long, afterwards, a 10~ms delay is added to ensure that the mechanical shutter has fully closed. The pump power $\approx1$~mW is three orders of magnitude greater than the ground-state PL saturation power, with typical photon energy $\approx5-10$~meV above the $X^{0}$ PL energy. To ensure the QD is unoccupied during the pump process, a large reverse bias is applied $V_{\rm{Gate}}=-2.4$~V.   

\subsection{Nuclear magnetic resonance (NMR)}

A copper wire coil is used to generate the magnetic field $B_{x}\perp z$ needed to carry out NMR experiments. The coil is positioned $\approx0.5$~mm from the QD sample. The coil is made of 10 turns of a 0.1~mm diameter enamelled copper wire wound on a $\approx0.4$~mm diameter spool in 5 layers, with 2 turns in each layer. A class-AB radiofrequency (Rf) amplifier (Tomco BT01000-AlphaSA rated up to 1000~W) is used to drive the coil, fed by the output of an arbitrary waveform generator (Keysight M8190).

The timing of NMR experiments proceeds as shown in Fig.~\ref{SupFig2:MmtCycle}. First, two Rf pulses are used to increase the NMR signal by transferring the nuclear spin population into the optimal configuration \cite{Dyte_2024,Chekhovich_2015}. Consider the case where the dot is pumped with $\sigma^{+}$ light leaving the $I_{\rm{z}}=-3/2$ nuclear spin states the most populated \cite{MillingtonHotze_2024}, while the $I_{\rm{z}}=+3/2$ states are nearly unpopulated. By maximising the population change we maximize the NMR signal. For measurements performed on the central transition the population difference between the $I_{\rm{z}}=-1/2$ and $I_{\rm{z}}=+1/2$ states needs to be maximized. To this end we apply two $\pi$ pulses to transfer the nuclear spin populations. First, a pulse is applied to the $+1/2\leftrightarrow+3/2$ transition, exchanging their populations leaving $I_{\rm{z}}=+1/2$ as the least populated state. Then, after a short delay, a second pulse is applied to the $-3/2\leftrightarrow-1/2$ transition leaving $I_{\rm{z}}=-1/2$ as the most populated state. A similar process can be applied in the case one of the satellite transitions is being investigated.

\begin{figure}
    \centering
    \includegraphics[width=0.98\linewidth]{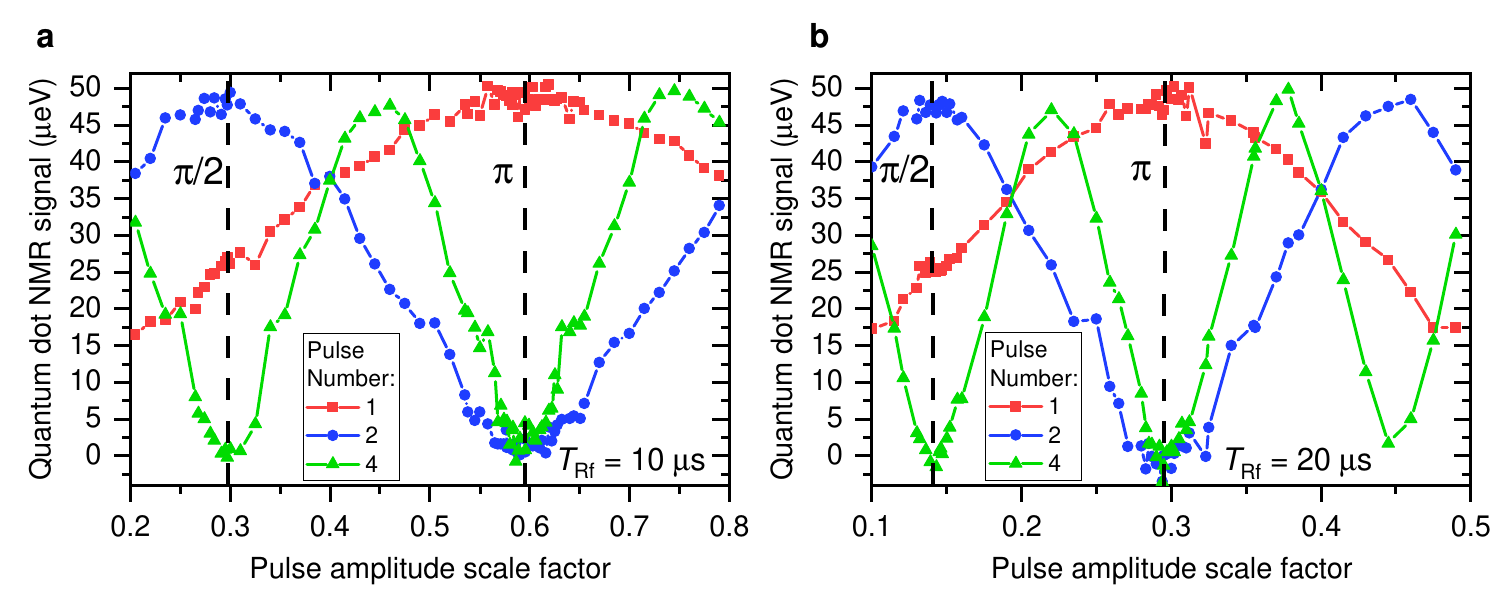}
    \caption{\textbf{a} Rabi oscillations of the nuclear spins under Rf pulse bursts of increasing amplitude with fixed pulse duration $T_{\rm{Rf}}=10~\mu$s. Increasing the pulse amplitude increases the degree of rotation of the spins. The first peak in the oscillation (solid squares, red, one pulse) corresponds to a $\pi$ rotation. We obtain the required amplitude of a $\pi$/2 rotation from the first peak for the oscillation driven by two pulses (solid circles, blue). Additionally we conduct measurements with 4, 8 and 16 pulses to improve the accuracy of the acquired amplitudes. Only the 4 pulse measurement (solid triangles, green) is shown for clarity. \textbf{b} Same as \textbf{a} but for pulses with $T_{\rm{Rf}}=20~\mu$s.}
    \label{SupFig:RabiOscillations}
\end{figure}

Next, the optically generated nuclear spin polarization needs to be converted from longitudinal polarization to transverse polarization. This is so the spins begin to decohere and the coherence time can be measured. We achieved this simply by using a single $\pi$/2 Rf pulse. The phase of the pulse can be varied to allow control over the initial orientation of the nuclear spins in the $xy$ plane. Dynamical decoupling is then applied. The pulses for a given sequence are applied in a way that they are equally spaced by time $\tau$, this means that while $T_{\rm{FreeEvol}}$ is increased the time $\tau$ between each pair of adjacent pulses increases uniformly. Due to hardware limitations, the minimal interpulse delay achievable is $\tau\approx 0.3$~$\mu$s. This places a lower limit on the free evolution time that could be measured for a given dynamical decoupling sequence, based on the sequences number of pulses. Finally, a $\pi$/2 pulse is used to rotate the nuclei back along the \textit{z} axis for optical readout. Again, the phase of finalization pulse can be varied. If the phase of the finalization pulse is the same as the initialization pulse, the nuclei are rotated such that their populations are antiparrallel compared to before the initialization pulse. Comparatively, if the finalization pulse is $\pi$ out of phase with the initialization pulse, then the nuclear spins are aligned parallel to the orientation before initialization. All measurements use initialization and finalization pulses with $\pi$ phase difference, as this allows us to avoid the range of nuclear polarizations where electron-nuclear bistability \cite{Braun_2006} accelerates the nuclear spin dynamics.

Before optical readout we again transfer the spin populations, this allows us to multiply the NMR signal by exploiting the entire $I=3/2$ Hilbert space. The process is the reverse of the process described above. First, a $\pi$ pulse is applied to the $-3/2\leftrightarrow-1/2$ transition followed by another $\pi$ pulse on the $+1/2\leftrightarrow+3/2$ transition. The reason for this is that after application of the dynamical decoupling sequence, the remaining nuclear spin polarization is encoded in the $-1/2\leftrightarrow+1/2$ states. Assuming a $\pi$ phase difference between initial and final pulses, the $I_{\rm{z}}=-1/2$ state will be most populated, while the $I_{\rm{z}}=+1/2$ state will be least populated. The reverse population transfer means the $I_{\rm{z}}=-3/2$ and $I_{\rm{z}}=+3/2$ become the most and least populated states respectively, encoding the NMR signal into the $I_{\rm{z}}=\pm3/2$ subspace, approximately tripling the NMR signal measured optically. During NMR a bias of $V_{\rm{Gate}}=-1.5$~V is applied to the sample to ensure that there are no resident electrons.

\subsection{Rf pulse calibration}

Dynamical decoupling requires that precise rotations of the nuclei are performed. For some of the measurements in this work, as many as 2400 Rf pulses were applied to drive rotations of the nuclear spins, any inaccuracy in the rotations would therefore worsen the coherence of the nuclear spin polarization. To this end, we performed Rabi oscillation measurements to calibrate the $\pi$ and $\pi$/2 Rf pulses. The Rf pulses are produced by modulating the Rf carrier with a raised cosine envelope function of total duration $T_{\rm{Rf}}$ between the two zero-amplitude points. For a desired $T_{\rm{Rf}}$, the amplitude of the applied pulses was increased and the resulting NMR signal measured, Supplementary Fig.~\ref{SupFig:RabiOscillations}a shows the Rabi oscillations observed for $T_{\rm{Rf}} = 10~\mu$s pulse duration. From this the power required for a $\pi$ pulse is obtained by locating the first peak in the oscillation. The power needed for a $\pi$/2 Rf pulse is found by applying two separate pulses. Here, the first peak now gives the amplitude needed for a $\pi$/2 pulse, as this corresponds to a total rotation of nuclei by $\pi$ meaning each of the two pulses will have rotated the nuclei by $\pi$/2. Finally, to improve the accuracy of these calibrations we conducted measurements with an increasing number of pulses: 4, 8, and 16. The Rabi oscillations were collectively fitted to obtain the power scaling factors. Supplementary Fig.~\ref{SupFig:RabiOscillations}b shows the same calibrations but for $T_{\rm{Rf}}=20~\mu$s pulses.

\begin{figure}
    \centering
    \includegraphics[width=0.7\linewidth]{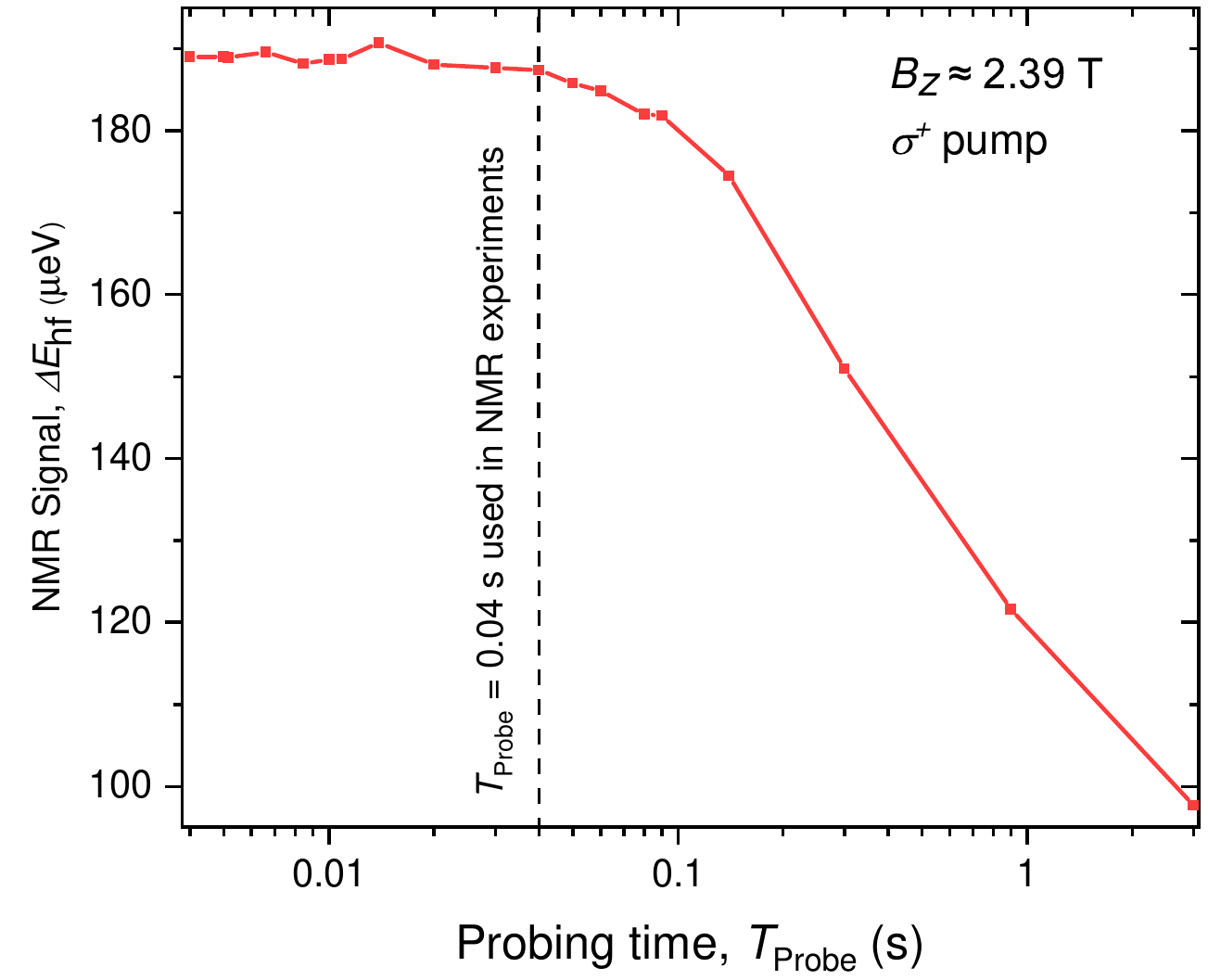}
    \caption{Calibration measurement for the optical probe duration. Hyperfine shift measured under different duration optical probing after being polarized with $\sigma^{+}$ optical pumping. The time ($T_{\rm{Probe}}$) used for coherence measurements and nuclear spin decoupling is shown by the dashed line. Measurement was conducted with external magnetic field $B_{\rm{z}}=2.39$~T.}
    \label{SupFig4:ProbeTime}
\end{figure}

\subsection{Optical probing of nuclear spins}

Measurement of the nuclear spin polarization is conducted using optical probing. The probe power and the bias during the probe ($V_{\rm{Gate}}=+0.9$~V) are chosen to maximize (saturate) the intensity of the ground neutral exciton (X$^{0}$) state PL. There is a 10~ms delay between the bias switching and the mechanical shutter activating, forming the probe pulse. This delay ensures that the charge state of the quantum dot is in its steady state when the optical probe pulse is applied. An example PL spectra is shown in Fig.~2a of the main text. The PL is generated by recombination of the electron-hole pairs. Under optical excitation the spin of the electron is random, meaning light is emitted from both bright exciton spin states. The PL is accumulated over millisecond timescales allowing both exciton states to be observed in the same PL spectrum. The applied magnetic field splits the two PL lines, these can be further shifted by the nuclear polarization via the hyperfine interaction. The shifts can be seen when the nuclei are polarized with different circularly polarized light. Injection of optically excited electrons into the dot leads to spin flip-flops between the electron spin and nuclei which will progressively depolarize the nuclei. A probe pulse duration measurement is shown in Supplementary Fig.~\ref{SupFig4:ProbeTime} for external magnetic field of $B_{\rm{z}}=2.39$~T. As the probe time is increased the nuclear polarization decreases, seen in the reduction of PL splitting. Similar results are observed in a wide range of magnetic fields $B_{\rm{z}}=1 - 8$~T. Based on such calibration results we use a probe time of $T_{\rm{Probe}}=0.04$~s maximizing the collected signal from the optical excitation without a significant loss in nuclear spin polarization due to spin flip-flops.

\section{Additional experimental results}

\subsection{Optimal Rf pulse duration in dynamical decoupling}
\begin{figure}
    \centering
    \includegraphics[width=0.9\linewidth]{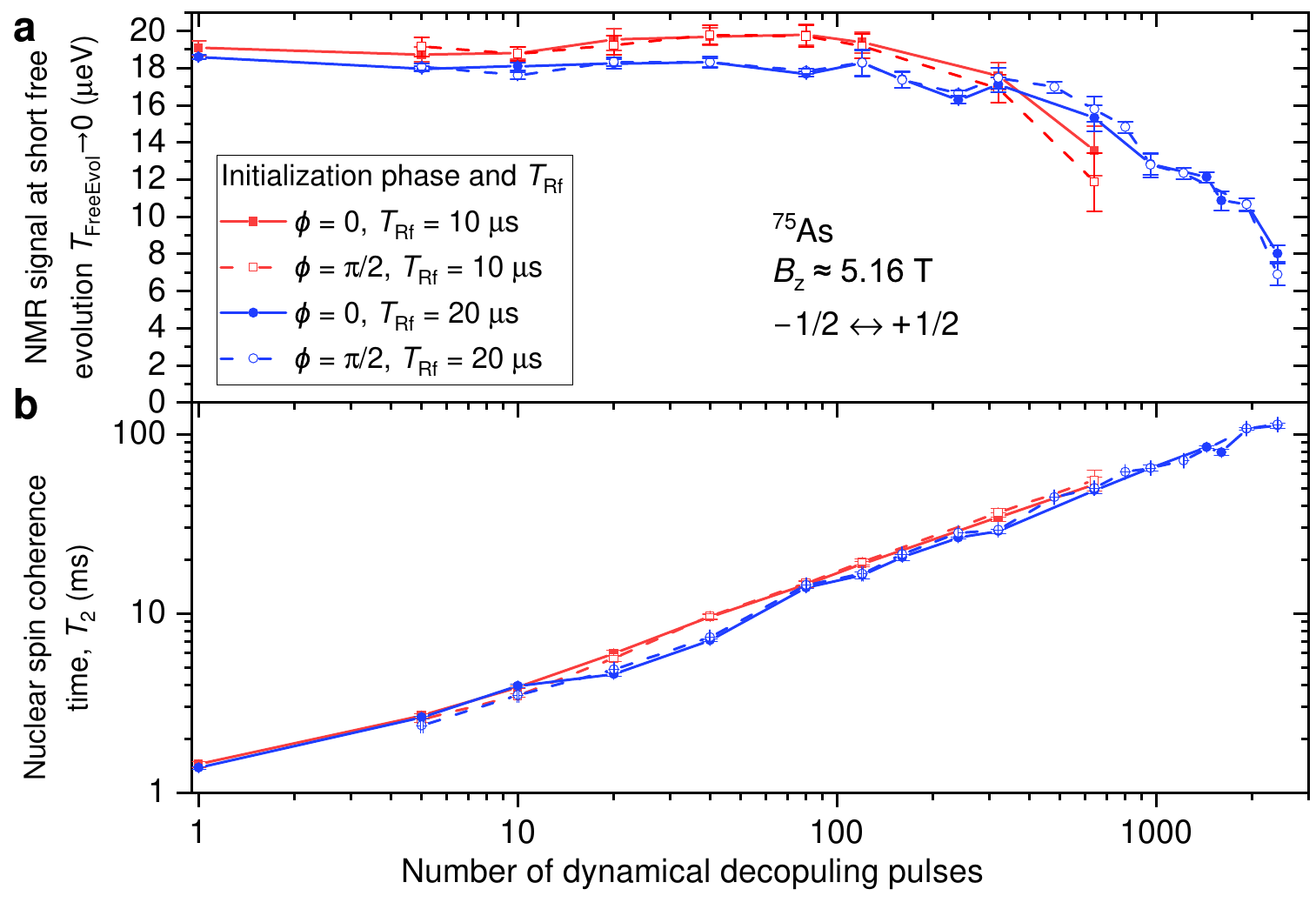}
    \caption{\textbf{a} NMR signal (spin echo amplitude) at short free evolution time $T_{\rm{FreeEvol}}\rightarrow 0$ plotted as a function of the number of Rf pulses in the dynamical decoupling sequence. The data is shown for Rf pulse durations of $T_{\rm{Rf}}=10~\mu$s (squares) and $T_{\rm{Rf}}=20~\mu$s (circles). Measurements are conducted with initial nuclear polarization along the $-y$ axis (solid symbols, initialization Rf pulse phase $\phi=0$) and the $x$ axis ($\phi=\pi/2$, open symbols) of the rotating frame. \textbf{b} Dependence of the nuclear spin coherence time $T_{2}$ on the number of Rf pulses.}
    \label{SupFig:Pulsetime}
\end{figure}

Ideally, nuclear spin rotations would be carried out instantaneously, minimising decoherence due to unwanted evolution of the nuclear spin ensemble during the rotation. As we are limited to using finite time Rf pulses, one may expect that shorter $T_{\rm{Rf}}$ times would yield better coherence storage performance. Yet, as discussed in the main text, we find that when using short $T_{\rm{Rf}}=10~\mu$s pulses we see a drop in performance for high pulse number sequences ($\approx 640$ pulses) compared to when longer $T_{\rm{Rf}}=20~\mu$s pulses are used. Here were present experimental results that reveal these counterintuitive observations. Nuclear spin decoherence measurements, such as shown in Fig.~2c of the main text, have been conducted both with $T_{\rm{Rf}}=10~\mu$s and $T_{\rm{Rf}}=20~\mu$s on the $I_{\rm{z}}=\pm 1/2$ nuclear spin subspace. In each such measurement, the total free evolution time $T_{\rm{FreeEvol}}$ is varied while keeping a constant number of pulses in the dynamical decoupling sequence. The resulting dependence of the spin echo NMR signal is fitted with a stretched exponential function of $T_{\rm{FreeEvol}}$. From the fit we derive the NMR signal in the limit of short free evolution $T_{\rm{FreeEvol}}\rightarrow 0$ and the characteristic nuclear spin coherence time $T_2$. These results are shown in Figs.~\ref{SupFig:Pulsetime}a and b, respectively. The values are shown as a function of the total number of pulses in Hahn echo (1 pulse), CHASE-5 (5 pulses), CHASE-10 (10 pulses), CHASE-20 (20 pulses), and a varying number of CHASE-40 cycles (all points with $\geq 40$ pulses). The results are shown for two different orientations of the initial nuclear spin polarization in the equatorial plane of the rotating frame (initialization pulse phase $\phi=0$ and $\phi=\pi/2$).

From Supplementary Fig.~\ref{SupFig:Pulsetime}a we find that for short pulse sequences ($\lesssim 100$ pulses) the short-evolution NMR signal is somewhat better (larger) for short pulses ($T_{\rm{Rf}}=10~\mu$s, squares), as these are better approximation to instantaneous spin Hamiltonian transformations than the $T_{\rm{Rf}}=20~\mu$s pulses (circles) pulses. However, for longer sequences (640 pulses) the short-evolution spin echo under short pulses ($T_{\rm{Rf}}=10~\mu$s) becomes worse than under long pulses ($T_{\rm{Rf}}=20~\mu$s).  
 Supplementary Fig.~\ref{SupFig:Pulsetime}b shows that the coherence time $T_{2}$, which is a measure of decoherence during free evolution, is nearly independent of the Rf pulse duration $T_{\rm{Rf}}$. This allows us to conclude that the reduced short-evolution NMR signal (i.e. reduced fidelity of coherence storage) at $T_{\rm{Rf}}=10~\mu$s is a result of undesired spin evolution during the pulses, rather than pure decoherence during free evolution between the pulses. In particular, we attribute the reduced performance of the $T_{\rm{Rf}}=10~\mu$s pulses to parasitic leakage of the stored state to the two nuclear satellite transitions, owing to the broader spectral profile of the pulses, as shown by the dashed lines in Fig.~1a of the main text.

\subsection{Comparison of pulsed spin locking in CHASE-20 and CHASE-40 decoupling sequences}

\begin{figure}
    \centering
    \includegraphics[width=0.9\linewidth]{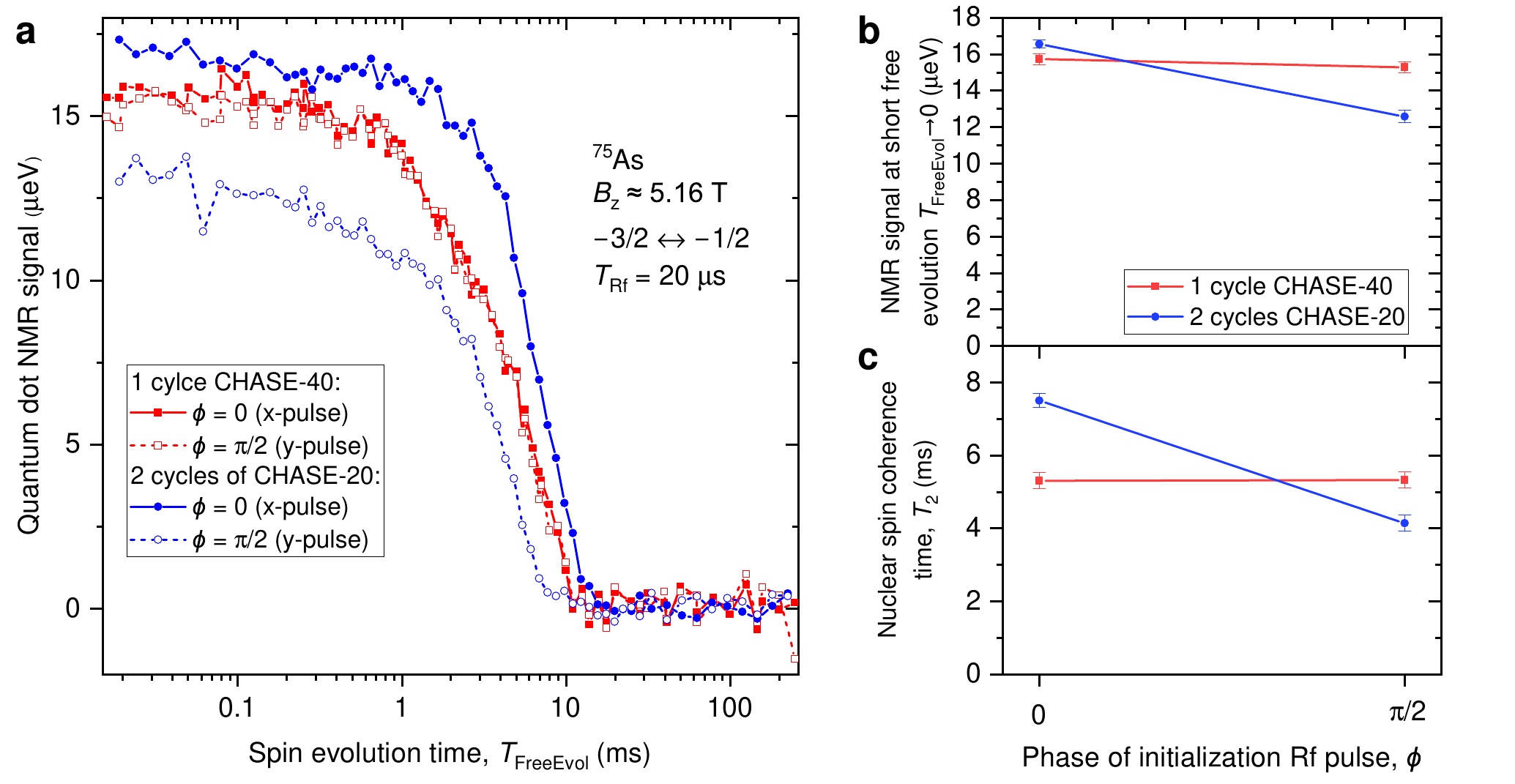}
    \caption{\textbf{a} Decoherence measured on the $I_{\rm{z}}=(-3/2,-1/2)$ nuclear spin subspace. Two cycles of CHASE-20 (squares) are compared to a single CHASE-40 cycle (circles) both when initializing the nuclear spin polarization along the $-y$ axis of the rotating frame (solid symbols, $\phi=0$ Rf pulse phase) and along the $x$ axis (open symbols, $\phi=\pi/2$). \textbf{b} NMR signal (spin echo amplitude) at short free evolution time $T_{\rm{FreeEvol}}\rightarrow 0$ plotted as a function of the phase $\phi$ of the initialization Rf pulse for two cycles of CHASE-20 (squares) and one cycle of CHASE-40 (circles). The values are obtained from fitting the decay curves in (\textbf{a}). \textbf{c} Same as (\textbf{b}) but for nuclear spin decoherence time T$_{2}$.}
    \label{SupFig:SpinLocking}
\end{figure}

Pulsed spin locking is observed in various dynamical decoupling techniques. The simplest example is the Carr-Purcell sequence, where a train of $\pi$ Rf pulses of the same phase is applied. The spin locking can be understood intuitively as an effective transverse magnetic field in the rotating frame that arises from repeated rotations of the nuclear spins around a preferential equatorial axis of the rotating frame \cite{Li_2008}. If the coherent nuclear polarization, created by the initial $\pi/2$ Rf pulse, is parallel to this effective transverse magnetic field, the nuclear state becomes ``locked'' and its decay is suppressed. In other words, the transverse ($T_2$) coherence decay is replaced by a decay that is akin to longitudinal ($T_1$) relaxation in the rotating frame. By contrast, the initial state polarized orthogonal to the effective locking field undergoes precession and accelerated decoherence. In the case of the CHASE-10 cycle, shown in Fig.~1b of the main text, the two $\pi$ rotations (labeled $+{\rm{x}}^2$ and $-{\rm{x}}^2$) create a preferential direction in the rotating frame, whereas the four $\pi/2$ rotation pulses labeled as $\pm {\rm{x}}$ are balanced by four $\pm {\rm{y}}$ Rf pulses rotating the spins around the orthogonal axis in the rotating frame.

The CHASE-40 sequence is designed to be resilient against pulsed spin locking, by combining four CHASE-10 subsequences, where the phase of all Rf pulses is stepped in each block by $\pi/2$, as shown in Fig.~1c of the main text. Here we perform direct comparison of CHASE-40 with a sequences of pulses, where such phase stepping is omitted. To this end we use two cycles of CHASE-20 sequences, where each CHASE-20 block is a combination of CHASE-10 and its mirror copy, where all Rf pulses are applied in reverse \cite{Waeber_2019}. Thus, both the CHASE-40 and 2 cycles of CHASE-20 differ only by the phases of the Rf pulses. In the case of CHASE-20, all the $\pi$ rotations ($+{\rm{x}}^2$ and $-{\rm{x}}^2$) are around the same equatorial axis of the rotating frame. 

The experiments are performed on the $I_{\rm{z}}=(-3/2,-1/2)$ subspace of the $^{75}$As nuclear spins. The resulting decays of nuclear spin echo with the increasing free evolution time $T_{\rm{FreeEvol}}$ are shown in Supplementary Fig.~\ref{SupFig:SpinLocking}a. The decay curves are fitted with stretched exponentials to derive the NMR signal in the limit of short free evolution $T_{\rm{FreeEvol}}\rightarrow 0$ (shown in Supplementary Fig.~\ref{SupFig:SpinLocking}b) and the characteristic nuclear spin coherence time $T_2$ (shown in Supplementary Fig.~\ref{SupFig:SpinLocking}c). In the case of CHASE-40 (squares in Supplementary Fig.~\ref{SupFig:SpinLocking}a) the decoherence (characterized by $T_2\approx5.3$~ms) does not depend on the direction of the initial transverse nuclear spin polarization. This is also in agreement with the data of Fig.~4 of the main text obtained for the $I_{\rm{z}}={-1/2,+1/2}$ nuclear spin subspace. These results confirm that CHASE-40 effectively eliminates the pulsed spin locking.

For the two cycles of CHASE-20 (circles in Supplementary Fig.~\ref{SupFig:SpinLocking}a) we observe a pronounced difference in decoherence, depending on the initial nuclear spin state. For the spin state polarized along the $-y$ axis of the rotating frame (solid circles, $\phi=0$) the decay is slowed down ($T_{2}\approx7.5$~ms), revealing the pulsed spin locking effect. (With the convention used here, the pulse labeled as ``+x'' in the sequence definition of Figs.~1b, c of the main text corresponds to $\phi=0$. This pulse produces Rf magnetic field along the $x$ axis of the rotating frame and initializes nuclear polarization along the $-y$ axis of the rotating frame. For the ``$-$x'' pulse the Rf phase is $\phi=\pi$. The ``+y'' pulse corresponds to $\phi=\pi/2$ and initializes the nuclear spin state along the $x$ axis. The $\pm{\rm{x}}^2$ Rf pulses have the same phase as $\pm{\rm{x}}$, and so on).

For the case of CHASE-20 decoupling with an initial nuclear spin polarization along the $x$ axis ($\phi=0$), there is a pronounced reduction both in the nuclear spin echo amplitude at short free evolution $T_{\rm{FreeEvol}}\rightarrow 0$ and in $T_2\approx4.1$~ms, characterizing decoherence during free evolution. In the context of quantum memories, this can be interpreted as the worst case scenario, which is an inevitable side effect of the pulsed spin locking. This is why elimination of spin locking, such as achieved with CHASE-40, is important for operating nuclear spins as a quantum memory, which is generally expected to preserve an arbitrary coherent state.

\subsection{Details on Measurement of Nuclear Spin Transitions}

The NMR Spectra of the $^{75}$As nuclei used in Fig.~1a of the main text were obtained using two seperate measurment techniques. For the $-1/2\leftrightarrow+1/2$ central transition (blue in Fig.~1a) the spectrum was obtained from a Fourier transform of the measured free induction decay. The free induction decay was measured using $T_{\rm{Rf}}=10~\mu s$. The $-3/2\leftrightarrow-1/2$ and $+1/2\leftrightarrow+3/2$ satellite transitions (black in Fig.~1a) were measured using the inverse NMR measurement technique \cite{Chekhovich_2012}.

\section{Discussion of quantum memory applicability for quantum repeaters}
Quantum repeaters are a necessary component for the realization of long distance communication of quantum information. The implementation of such devices will enable the secure communication of information, paving the way for the quantum internet \cite{Zhao_2008,Tittel_2010,Azuma_2023}. The need for quantum repeaters stems from the limitations on information transfer distance in optical fibers. Photons are superb 'flying' qubits that can be used for long distance communication of quantum states owing to their long coherence times. Yet modern optical fibers can only be used to transmit photons for distances of $\approx 20-100$~km due to attenuation. One solution to this problem is to transmit the quantum information via satellites, removing the issue of losses due to fiber attenuation. Such systems have been demonstrated, enabling quantum teleportation over distances of up to 1400~km, yet issues such as sensitivity to atmospheric turbulence and the requirement for direct line of sight with the satellite remain \cite{Ren_2017}. Another alternative is to use quantum repeaters, which seek to get around this limitation by splitting the full communication distance into several shorter channels connected by quantum repeater nodes, by entangling all of the nodes, entanglement swapping can then be used to transmit information between the two end nodes over distances greater than those possible by direct photon transmission \cite{Zhao_2008,Tittel_2010,Azuma_2023,Briegel_1998}. This simplified approach requires that entanglement generation occurs on every repeater node simultaneously, which significantly limits the chance of success. By using a quantum memory qubit which possess a long coherence time $T_{2}$, the entangled states can be stored for time $T_{2}$ allowing multiple attempts at generating entanglement. Therefore, by using time multiplexing, the chance of successful entanglement swapping can be increased \cite{Azuma_2023,Duan_2001}. 

The storage time of the memory required to achieve the transfer of information over a useful distance varies significantly depending on the quantum repeater protocol. In particular, the required communication time can be reduced by introducing more complexity to the system, such as multiplexing \cite{Collins_2007,Jiang_2007}. Therefore the memory time needed for a $\approx~1000$~km channel can vary from $1000$~s down to $1$~ms \cite{Zhao_2008,Tittel_2010,Sharman_2021}. Fundamentally, the maximum distance achievable for a given memory time is tied to the speed of light, therefore memories of $1$~ms are sufficient to beat direct photon transmission \cite{Tittel_2010} while $5$~ms is sufficient to reach distances of $1000$~km, a commonly used benchmark in the literature \cite{Zhao_2008}. Therefore, using this approach, we estimate that a quantum memory based on nuclear spins with $T_{\rm{M}}\approx100$~ms would be able to store a given state for long enough to send a message over a distance greater than half the Earth's circumference, meaning it would be sufficient for world wide communication. However, photon transmission time is not the only limitation of a quantum repeater \cite{Sharman_2021,Langenfeld_2021}. Other limiting factors include: light-matter entanglement generation \cite{DeGreve_2012,Coste_2023,Laccotripes_2024}, state transfer to the nuclei \cite{Gangloff_2019,Appel_2024}, entanglement swapping \cite{Stockill_2017}, and state measurement \cite{Atature_2006}. While all these building blocks have been demonstrated for QDs, they would all consume time, reducing the amount of time for generating entanglement. Out of the factors listed above, the generation of entanglement is the main limitation on communication distance. In the above discussion we assumed a unity fidelity of entanglement generation. In reality this is not the case and the photon transmission process needs to be repeated multiple times until the entanglement is established, which significantly limits the achievable distances for quantum repeaters \cite{Yu_2020}. For QDs, entanglement generation rates as high as 7.3~kHz have been demonstrated \cite{Stockill_2017}. We can therefore estimate, using the equations and parameters reported in Ref. \cite{Yu_2020}, that with the current memory storage time of $T_{\rm{M}}\approx100$~ms it should be possible to achieve quantum repeater communication distances of $\approx 13.4$~km. Assuiming a link efficiency of 0.34, as used in \cite{Yu_2020} leads to a more favorable estimate of $\approx 39.4$~km, making nuclear spin based memories competitive with the atomic ensemble systems used in Ref. \cite{Yu_2020}, as well as matching communication distances achievable with direct fiber transmission of photons.

Although current nuclear spin memory times do not yet match those achievable in certain solid state or atomic systems \cite{Wang_2017,Wang_2021,Atatre_2018}, the III-V semiconductor quantum dots benefit from superior optical properties such as efficient generation of single \cite{Neuwirth_2021,Huber_2017,Schweickert_2018,Arakawa_2020} and entangled photons \cite{Liu_2019,Rota_2024}, which are also required for quantum communication. QDs also benefit from the well established semiconductor fabrication techniques, offering a path for scalability. The storage times demonstrated here for nuclear spins, combined with the high optical entanglement generation rates mean that QDs should be able to compete with other existing approaches to quantum repeaters \cite{Stockill_2017,Yu_2020}. Finally, by using multiplexing or alternatively, by extending the coherence time to $T_{2}=1$~s as discussed in the main text, it should be possible to extend the possible communication distance to $500-1000$~km, sufficient to outperform direct photon transmission and enable quantum repeater channels that, for example, connect major cities \cite{Collins_2007,Sharman_2021}.

\section{Average Hamiltonian theory applied to design of {{CHASE}} dynamical decoupling sequences}

\subsection{Nuclear spin interactions}

The Zeeman term accounts for the coupling of the QD nuclear spins ${\bf{I}}_k$ to the static magnetic field $B_{\rm{z}}$ directed along the $z$ axis. In the laboratory frame it can be written as
\begin{align}
    \mathcal{H}_{\rm{Z,N}} = -\sum_{k=1}^{N} \hbar\gamma_k B_{\rm{z}}\hat{I}_{{\rm{z}},k},\label{Eq:HZN}
\end{align}
where the summation goes over all individual nuclei $1\leq k \leq N$, $\hbar=h/(2\pi)$ is the reduced Planck's constant, $\gamma_{k}$ is the gyromagnetic ratio of the $k$-th nuclear spin and $\hat{\bf{I}}_k$ is a vector of spin operators with Cartesian components $(\hat{I}_{{\rm{x}},k},\hat{I}_{{\rm{y}},k},\hat{I}_{{\rm{z}},k})$. The result of the Zeeman term alone is a spectrum of equidistant single-spin eigenenergies $-I_{\rm{z}} \hbar \gamma_{k}B_{\rm{z}}$. These $2I+1$ states are also the eigenstates of the $\hat{I}_{{\rm{z}}}$ spin projection operator with eigenvalues $I_{\rm{z}}$ satisfying $-I\leq I_{\rm{z}}\leq +I$. 

The interaction of the nuclear electric quadrupolar moment with the electric field gradients is described by the term (Ch.~10 in Ref.~\cite{SlichterBook}):
\begin{align}
\mathcal{H}_{\rm{Q,N}} = \sum_{k=1}^N
\frac{q_{k}}{6}[3\hat{I}_{{\rm{z'}},k}^2-I_{k}^2+\eta_{k}(\hat{I}_{{\rm{x'}},k}^2-\hat{I}_{{\rm{y'}},k}^2)],\label{Eq:HQN}
\end{align}
where $q_{k}$ and $\eta_{k}$ describe the magnitude and asymmetry of the electric field gradient tensor, whose principal axes are $x'y'z'$. The strain is inhomogeneous within the QD volume, so that $q_{k}$ and $\eta_{k}$ vary between the individual nuclei. The axes $x'y'z'$ are different for each nucleus and generally do not coincide with crystallographic axes or magnetic field direction. For the as-grown (unstrained) GaAs/AlGaAs QDs the typical quadrupolar shift is around $\vert q_{k} \vert/h\approx24$~kHz for $^{75}$As \cite{Ulhaq_2016,MillingtonHotze_2023}. The full width at half maximum of the $\vert q_{k} \vert/h$ distribution is $\approx14~$kHz. For a small fraction of arsenic nuclei, adjacent to aluminium atoms, the shifts are as large as $\vert q_{k}\vert/h\approx200$~kHz \cite{Zaporski_2023}. In the strained sample structure, $q_{k}$ are dominated by the extrinsic uniaxial stress, with the typical values $\vert q_{k}\vert/h\approx250$~kHz for $^{75}$As in the present work. All experiments are conducted under sufficiently strong magnetic fields, where $|\hbar\gamma_k B_{\rm{z}}|\gg |q_{k}|$ and quadrupolar effects can be treated perturbatively. In this perturbative regime, the main effect of the quadrupolar shifts is the anharmonicity of the nuclear spin eigenenergies and the resulting quadrupolar NMR multiplet of $2I$ magnetic-dipole transitions, split by $\nu_{\rm{Q}}\approx q_{k}/h$. The $I_{\rm{z}}=\pm1/2$ projection states of a half-integer nuclear spin are influenced by quadrupolar effects only in the second order. These second order shifts scale as $\propto\nu_{\rm{Q}}^2/\nu_{\rm{N}}$, where $\nu_{\rm{N}}=\gamma B_{\rm{z}}/(2\pi)$ is the nuclear spin Larmor frequency.

Since we apply dynamical decoupling only to the two-level subspaces of the spin 3/2 four-level Hilbert space, we can use the standard rotating frame representation. For the $I_{\rm{z}}=\pm 1/2$ subspace the reference frequency of the frame is set to the Larmor frequency $\nu_{\rm{N}}$, while for the satellite subspaces the reference frequency is set to $\nu_{\rm{N}}\pm\nu_{\rm{Q}}$ to match the NMR frequency of the corresponding satellite transition. The rotating frame transformation in the following effective Zeeman Hamiltonian term
\begin{align}
    \mathcal{H}'_{\rm{Z,N}} = -\sum_{k=1}^{N} h \Delta \nu_k \hat{I}_{{\rm{z}},k},\label{Eq:HZNRot}
\end{align}
where $\Delta \nu_k$ is the resonance offset of the $k$th nucleus. In the case of the $I_{\rm{z}}=\pm 1/2$ spin subspace, $\Delta \nu_k$ represents the inhomogeneous second order quadrupolar shifts, which are within $\lesssim1$~kHz. For the $I_{\rm{z}}=(-3/2,-1/2)$ satellite transition subspace the set of $\Delta \nu_k$ describes the distribution of the first order quadrupolar shifts which arise from inhomogeneous strain within the QD volume and range from tens to hundreds of kHz for the individual $^{75}$As nuclei.

Direct interaction between the nuclei is described by the dipole-dipole Hamiltonian:
\begin{align}
\mathcal{H}_{\rm{DD}}=\sum_{1\leq j<k\leq N}b_{j,k}\left(3\hat{I}_{{\rm{z}},j}\hat{I}_{{\rm{z}},k}-\hat{\bf{I}}_j{\bf{\cdot}}\hat{\bf{I}}_k\right){\rm{,}}\nonumber\\
b_{j,k}=\frac{\mu_0 \hbar^2}{4\pi}\frac{\gamma_{j}\gamma_{k}}{2}\frac{1-3\cos^2{\theta_{j,k}}}{r_{j,k}^3}\label{Eq:HDD}
\end{align}
Here, $\mu_0=4\pi\times 10^{-7}\;{\rm{N\;A}}^{-2}$ is the magnetic constant and $r_{j,k}$ denotes the length of the vector, which forms an angle $\theta$ with the $z$ axis and connects the two spins $j$ and $k$. The Hamiltonian of Supplementary Eq.~\ref{Eq:HDD} has been truncated to eliminate all spin non-conserving terms -- this is justified for static magnetic field exceeding $\gtrsim1$~mT. The spin-spin interaction constants can be written in frequency units $\nu_{jk}=b_{j,k}/h$. The typical magnitude of the interaction constants for the nearby nuclei in GaAs is $\max{\vert \nu_{jk}\vert }\approx100$~Hz. Consequently, the typical timescales of the decoherence driven by the many-body dipole-dipole interactions are on the order of $T_2\approx 1$~ms. The dipole-dipole interaction does not change its form under rotating frame transformation.

\subsection{Average Hamiltonian theory}

The basics of average Hamiltonian theory can be found in relevant textbooks \cite{Mehring_1983}. Here we briefly outline the key points. The effect of the Rf pulses can be viewed as transformation of the many-body nuclear spin Hamiltonian. In the limit of short and strong Rf pulses, we can treat the Hamiltonian in a toggling reference frame as a piece-wise function of time, which remains constant in between the Rf pulses. After the first pulse, the Hamiltonian can be written as:
\begin{equation}
\mathcal{H}_1 = P_1^{-1} \mathcal{H}_0 P_1,\label{eq:Ht1}
\end{equation}
where $\mathcal{H}_0 = \mathcal{H}'_{\rm{Z,N}} + \mathcal{H}_{\rm{DD}}$ is the initial Hamiltonian in the rotating frame and $P_1$ is the unitary transformation operator. After the second and third Rf pulses, the Hamiltonian is:
\begin{eqnarray}
\mathcal{H}_2 = (P_2 P_1)^{-1} \mathcal{H}_0 P_2 P_1,\\
\mathcal{H}_3 = (P_3 P_2 P_1)^{-1} \mathcal{H}_0 P_3 P_2 P_1,\label{eq:Ht23}
\end{eqnarray}
and so on for the subsequent pulses. The evolution of the spin ensemble can in principle be calculated using the Hamiltonians, such as in equations \ref{eq:Ht1}, \ref{eq:Ht23}, to derive the unitary propagators in each time segment. However, for long pulse sequences this quickly becomes impractical. Instead, we want to find the propagator over the entire pulse sequence cycle $T_{\rm{Cycle}}$ 

\begin{eqnarray}
U(T_{\rm{Cycle}}) = \exp{(-i T_{\rm{Cycle}} \mathcal{H}_{\rm{eff}})},\label{eq:UAHT}
\end{eqnarray}
where $\mathcal{H}_{\rm{eff}}$ is some effective Hamiltonian describing spin interactions in the toggling frame over the pulse sequence cycle. Such an effective Hamiltonian can indeed be found in the form of a Magnus expansion
\begin{eqnarray}
\mathcal{H}_{\rm{eff}} =  \mathcal{H}^{(0)} + \mathcal{H}^{(1)} + \mathcal{H}^{(2)}...,\label{eq:HAHT}
\end{eqnarray}
where the first terms of the expansion are
\begin{align}
    \mathcal{H}^{(0)}&=\frac{1}{ T_{\rm{Cycle}} }\int_0^{T_{\rm{Cycle}}}\tilde{\mathcal{H}}(t)\mathrm{d} t\;\mathrm{,}\label{HMagnus0}\\
    \mathcal{H}^{(1)}&=\frac{-i}{2T_{\rm{Cycle}}}\int_0^{T_{\rm{Cycle}}}\mathrm{d} t_2\int_0^{t_2}\mathrm{d} t_1\left[\tilde{\mathcal{H}}(t_2),\tilde{\mathcal{H}}(t_1)\right]\;\mathrm{,}\label{HMagnus1}\\
    \begin{split}
        \mathcal{H}^{(2)}&=\frac{1}{6T_{\rm{Cycle}}}\int_0^{T_{\rm{Cycle}}}\mathrm{d} t_3\int_0^{t_3}\mathrm{d} t_2\int_0^{t_2}\mathrm{d} t_1\Big(\left[\tilde{\mathcal{H}}(t_1),\left[\tilde{\mathcal{H}}(t_2),\tilde{\mathcal{H}}(t_3)\right]\right]+\left[\tilde{\mathcal{H}}(t_3),\left[\tilde{\mathcal{H}}(t_2),\tilde{\mathcal{H}}(t_1)\right]\right]\Big)\;\mathrm{.}\label{HMagnus2}
    \end{split}
\end{align}

The $\mathcal{H}^{(0)}$ term is the average Hamiltonian over the pulse sequence cycle. In all CHASE sequences it is eliminated $\mathcal{H}^{(0)}=0$ in the limit of short Rf control pulses $T_{\rm{RF}}\rightarrow0$. For the term $\mathcal{H}^{(n)}$, the integration volume scales as $\propto T_{\rm{Cycle}}^{n+1}$. Thus, as long as $\mathcal{H}^{(0)}=0$, one can ensure the convergence of the effective Hamiltonian to zero $\mathcal{H}_{\rm{eff}}\rightarrow 0$ in the limit of a short cycle $T_{\rm{Cycle}}\rightarrow 0$ (i.e. in the limit of fast dynamical decoupling). This is the main goal of the time-suspension dynamical decoupling. As a general rule, one seeks to eliminate the lowest order terms in the expansion, in order to achieve faster convergence of the effective Hamiltonian.

\subsection{Average Hamiltonians of the CHASE cycles}

We have calculated average Hamiltonian terms of the CHASE sequences up to the second order in Magnus expansion, inclusive. We take into account the finite (nonzero duration, $T_{\rm{Rf}}>0$) Rf pulses, which represent the time segments within the sequence cycle where the toggling frame Hamiltonian depends explicitly on time. For simplicity, we only treat Rf pulses with rectangular envelope function. Although this is different from the raised cosine pulses used in experiment, the rectangular pulses are sufficient to derive the main effect of the finite control pulses on the average Hamiltonian terms.

We start with the CHASE-10 sequence cycle. The leading term appears in the average Hamiltonian $\mathcal{H}^{(0)}$ under finite pulses $T_{\rm{Rf}}>0$. This Hamiltonian term has the form of a dipole-dipole interaction (Equation \ref{Eq:HDD}) but with an effective quantization axis along the $y$ direction (i.e. $3\hat{I}_{{\rm{z}},j}\hat{I}_{{\rm{z}},k}$ is replaced with $3\hat{I}_{{\rm{y}},j}\hat{I}_{{\rm{y}},k}$):
\begin{equation}
\mathcal{H}^{(0)}_{\rm{CHASE-10}} = h\frac{T_{\rm{Rf}}}{T_{\rm{Cycle}}}\sum_{i<j}\nu_{ij}(3I_{i,y}I_{j,y}-\textbf{I}_{i}\cdot\textbf{I}_{j}).\label{eq:H0CHASE10}
\end{equation}
This interaction causes preferential spin locking of the nuclear spin states polarized along the $y$ axis of the rotating frame. There are also non-zero first and second order terms in the Magnus expansion of the CHASE-10 effective Hamiltonian. The exact expressions can be derived, but are bulky. These higher order terms can be neglected, since they appear only for finite Rf pulses $T_{\rm{Rf}}>0$ and are dominated by the zero order term $\mathcal{H}^{(0)}$ in the practically useful limit of a short cycle time $T_{\rm{Cycle}}$.

The expression for the dominant zero order term of CHASE-20 is twice that of CHASE-10 (Eq.~\ref{eq:H0CHASE10}). However, since $T_{\rm{Cycle}}$ appears in denominator of Eq.~\ref{eq:H0CHASE10} and is twice longer for CHASE-20, the resulting zero order average Hamiltonian is the same as for CHASE-10. CHASE-20 is a symmetrised version of CHASE-10, i.e. it consists of a CHASE-10 subcycle and second CHASE-10 subcycle where all pulses are applied in reverse order. Symmetrisation leads to cancellation of the first order Magnus term, as well as all other terms with odd orders.

We now consider CHASE-40, the new sequence implemented in this work. The zero order (average) Hamiltonian term appears only under finite pulses $T_{\rm{Rf}}>0$ and has the following form
\begin{equation}
\mathcal{H}^{(0)}_{\rm{CHASE-40}} = -2h\frac{T_{\rm{Rf}}}{T_{\rm{Cycle}}}\sum_{i<j}\nu_{ij}(3I_{i,z}I_{j,z}-\textbf{I}_{i}\cdot\textbf{I}_{j})\label{eq:H0CHASE40}
\end{equation}
Compared to Eq.~\ref{eq:H0CHASE10}, there is an additional factor of 2, but $T_{\rm{Cycle}}$ in the denominator is 4 times larger. Thus the $\mathcal{H}^{(0)}_{\rm{CHASE-40}}$ term is twice smaller in magnitude compared to $\mathcal{H}^{(0)}_{\rm{CHASE-10}}$. Moreover, the symmetry axis of $\mathcal{H}^{(0)}_{\rm{CHASE-40}}$ coincides with the external magnetic field ($z$ axis). Thus, there is no preferential direction in the equatorial plane of the rotating frame, explaining the absence of spin locking.

CHASE-40 is not symmetric, so the first order Magnus term is not canceled. However, for all physically relevant parameters the first order term is found to be dominated, either by the zero order term (at short $T_{\rm{Cycle}}$) or by the second order term discussed below (for long $T_{\rm{Cycle}}$). The small first order term is not a major issue since it can be easily eliminated with an 80-pulse symmetrised supercycle \cite{Burum1979,Mehring_1983}.

The second order term of CHASE-40 is present even for ideal Rf pulses $T_{\rm{Rf}}\rightarrow0$. Thus we only consider this limit of delta-pulses and ignore all the corrections arising from finite pulses $T_{\rm{Rf}}>0$. Even then, the expression for the second order term of CHASE-40 is rather bulky. There are different types of contributions. In particular there are the so called cross terms which appear only in presence of both the resonance offset of the individual spins ($\Delta\nu_{k}\neq 0$) and the dipole-dipole interactions between the spins ($\nu_{jk}\neq 0$). These cross terms play an important role for the $I_{\rm{z}}=(-3/2,-1/2)$ spin subspace subject to a considerable first order quadrupolar broadening, leading to $\Delta\nu_{k}$ on the order of tens of kHz. For the homogeneous $I_{\rm{z}}=\pm1/2$ subspace, affected only by the second order quadrupolar shifts, the resonance offsets $\Delta\nu_{k}$, and hence the cross term in $\mathcal{H}^{(2)}_{\rm{CHASE-40}}$, can be made arbitrarily small, for example by increasing the static magnetic field, which increases the nuclear Larmor frequency $\nu_{\rm{N}}$. But even in the absence of any resonance offsets $\Delta\nu_{k}=0$, there remains a second order Magnus term of the following form
\begin{align}
\mathcal{H}^{(2)}_{\rm{CHASE-40}} = h\frac{T_{\rm{Cycle}}^2}{110592}\sum_{i<j}\Big[\big(3\sum_{k\neq i,j}(\nu_{ik}+\nu_{jk})(\nu_{ik}\nu_{jk}+\nu_{ij}\nu_{ik}+\nu_{ij}\nu_{jk})\big)+\nonumber\\ 
+8\nu_{ij}(\Delta\nu_{i}^2+\Delta\nu_{i}\Delta\nu_{j}+\Delta\nu_{j}^2)\Big](3I_{i,z}I_{j,z}-\textbf{I}_{i}\cdot\textbf{I}_{j})\label{eq:H2CHASE40}
\end{align}
This term has a form of an effective dipolar coupling between spins $i$ and $j$, summed over all unique pairs $i<j$. Such interaction can arise from the direct dipole-dipole interaction of spins $i$ and $j$ in presence of a resonance offset for at least one of the spins from the pair (the $8\nu_{ij}(\Delta\nu_{i}^2+\Delta\nu_{i}\Delta\nu_{j}+\Delta\nu_{j}^2)$ term). However, the interaction remains even if all $\Delta\nu_{k}=0$, but requires dipolar coupling to some third spin $k$. Such interaction can be interpreted as a three particle effect, where the effective interaction of spins $i$ and $j$ is mediated by all other spins $k\neq i,j$. There are some further second order terms that appear only under non-zero resonance offsets, but these are small and are omitted in Eq.~\ref{eq:H2CHASE40}.

The $\mathcal{H}^{(2)}_{\rm{CHASE-40}}$ term of Eq.~\ref{eq:H2CHASE40} dominates the residual toggling-frame Hamiltonian of CHASE-40 in the limit of large $T_{\rm{Cycle}}$. It is responsible for the quadratic growth of the decoherence rate $T_{\rm{M}}^{-1}$ with increasing $T_{\rm{Cycle}}$ at large $T_{\rm{Cycle}}\gtrsim2$~ms, as observed in Fig.~3e of the main text for the measurement on the CT subspace $I_{\rm{z}}=\pm1/2$. The zero order term $\mathcal{H}^{(0)}_{\rm{CHASE-40}}$ (Eq.~\ref{eq:H0CHASE40}) is negligible at large $T_{\rm{Cycle}}$ due to the small ratio ${T_{\rm{Rf}}}/{T_{\rm{Cycle}}}$, but becomes dominant in the limit of fast dynamical decoupling (when most of the sequence cycle is taken up by the control pulses, corresponding to ${T_{\rm{Rf}}}/{T_{\rm{Cycle}}}\approx 1/40$). The zero order term $\mathcal{H}^{(0)}$ is responsible for the growth of the decoherence rate $T_{\rm{M}}^{-1}$ with reducing $T_{\rm{Cycle}}$ at small $T_{\rm{Cycle}}\gtrsim2$~ms, as observed in Fig.~3e of the main text for the measurement on the CT subspace $I_{\rm{z}}=\pm1/2$. The slowest decoherence (the longest spin memory time $T_{\rm{M}}$) is achieved when the contributions of the zero and second order terms are comparable. This observation allows us to estimate the maximum $T_{\rm{M}}$ that would be achieved if the effect of finite pulses was eliminated \cite{Haeberlen_1968, Burum1979}. Assuming the second order term would be unaffected, we can extrapolate the measured power-law dependence of $T_{\rm{M}}$ on $T_{\rm{Cycle}}$ at large $T_{\rm{Cycle}}\gtrsim2$~ms into the range of small $T_{\rm{Cycle}}$. This yields an expected $T_{\rm{M}}\approx 1$~s for the shortest possible $T_{\rm{Cycle}}=0.8$~ms of a CHASE-40 cycle at $T_{\rm{RF}}=20~\mu$s used in this work. This prediction shows that another order of mangitude improvement in the quantum memory storage time might be within reach with the existing material parameters (such as strain) and using only pulse sequence design techniques \cite{Haeberlen_1968, Burum1979}.

Even once the effective Hamiltonian (Eq.~\ref{eq:HAHT}) is known, finding the spin dynamics is still a difficult problem. Formally, the relaxation function of the transverse nuclear spin polarization can be written as an infinite power series over time, where the coefficients are the moments of the NMR spectral lineshape (Ch. 6 in \cite{CowanBook}). In practice, only the second ($M_2$) and the fourth ($M_4$) moments can be calculated \cite{VanVleck_1984}, making the power series rather inaccurate. It is more practical to approximate the lineshape with a function, such as Gaussian, and use the second moment $M_2$ as a linewidth parameter. The approximate coherent memory time can then be found as 
\begin{align}
T_{\rm{M}}\approx\sqrt{2/M_2}.
\label{eq:T2M2}
\end{align}
For nuclear spin polarization along the $x$ axis of the rotating frame, the second moment can be caclualted as 
\begin{align}
M_2 = -{\rm{Tr}}\left\{[\mathcal{H}_{\rm{eff}},\sum_{k=1}^{N}\hat{I}_{{\rm{x}},k}]^2\right\}/{\rm{Tr}}\left\{\left(\sum_{k=1}^{N}\hat{I}_{{\rm{x}},k}\right)^2\right\}
\label{eq:M2}
\end{align}
The second moments can be calculated for an arbitrary orientation of the nuclear spin polarization by substituting the corresponding spin operators (e.g. $\hat{I}_{{\rm{y}},k}$) in Eq.~\ref{eq:M2}.

For practical calculations, Eq.~\ref{eq:M2} is applied to a system of 3 nuclear spins, assuming the same dipolar coupling $\nu_{\rm{DD}}$ between each spin pair, and the resonance offsets of $0, +\Delta\nu, -\Delta\nu$. The effective Hamiltonians are calculated for free induction decay (FID, no dynamical decoupling), Hahn echo (single $\pi$ refocusing pulse), and CHASE-40, and are substituted into Eqns.~\ref{eq:T2M2}, \ref{eq:M2} to derive the spin memory times. The values for FID and Hahn echo are equated to the measured $T_2$ values (the difference between $T_2$ and $T_{\rm{M}}$ is negligible if there is at most one refocusing pulse). This allows the coupling parameters $\nu_{\rm{DD}}$ and $\Delta\nu$ to be extracted and then used to calculate the spin memory time $T_{\rm{M}}$ for the CHASE-40 sequence. The resulting $T_{\rm{M}}^{\rm{CHASE-40}}$ is shown by the dashed lines as a function of $T_{\rm{Cycle}}$ in Fig.~3e of the main text and is discussed therein.

\section{Numerical modeling of nuclear spin decoherence and dynamical decoupling}

We perform exact numerical modeling on an ensemble of $N$ spin-1/2 nuclei coupled through dipole-dipole interactions (Supplementary Eq.~\ref{Eq:HDD}). The Zeeman effect of the strong static magnetic field is eliminated through rotating frame transformation. This leaves the Zeeman offset Hamiltonian (Supplementary Eq.~\ref{Eq:HZNRot}), where $\Delta \nu_k$ is the resonance frequency shift of the $k$-th nucleus, which includes the inhomogeneous nuclear qaudrupolar effects and chemical shifts.

During the Rf pulses, the time-dependent Hamiltonian is added:
\begin{eqnarray}
\mathcal{H}_{\rm{rf,N}} =-h\sum_{k} \nu_{1}(t) \left(\cos(\phi)\hat{I}_{{\rm{x}},k}-\sin(\phi)\hat{I}_{{\rm{y}},k}\right),\label{Eq:HRF}
\end{eqnarray}
where the summation goes over all nuclei, $\nu_{1}(t)$ is the slowly-varying Rf pulse envelope, and $\phi$ describes the phase of the Rf carrier and the corresponding orientation of the transverse magnetic field in the equatorial plane of the rotating frame. The envelope function has a raised cosine profile $\nu_{1}(t)\propto (1-\cos(2\pi (t-t_0) / T_{\rm{Rf}}))/2$, where $t_0$ is the starting time of the Rf pulse burst.

Most of the numerical modeling is carried out for a system of $N= 12$ $^{75}$As nuclear spins. The $\{x,y,z\}$ coordinates of the nuclei in units of nm are as follows:
\begin{align}
&\{0.,0.,0.\},\nonumber\\
&\{0.282393,0.282393,0.\},\nonumber\\
&\{0.282393,0.,0.282393\},\nonumber\\
&\{0.282393,-0.282393,0.\},\nonumber\\
&\{0.282393,0.,-0.282393\},\nonumber\\
&\{0.,0.282393,0.282393\},\nonumber\\
&\{-0.282393,0.282393,0.\},\nonumber\\
&\{0.,0.282393,-0.282393\},\nonumber\\
&\{-0.282393,0.,0.282393\},\nonumber\\
&\{0.,-0.282393,0.282393\},\nonumber\\
&\{-0.282393,-0.282393,0.\},\nonumber\\
&\{-0.282393,0.,-0.282393\}.
\end{align}

The unitary evolution of the system is simulated through numerical propagation of the Schrödinger equation from an initial wave function state $\psi_{\rm{Init}}$. The computation is carried out using the software package Wolfram Mathematica 13.2 and the Python QuTiP 4.7.3 package \cite{JOHANSSON_2013}. We chose $\psi_{\rm{Init}}$ as eigenstates of the time-independent Hamiltonian, which is a sum of Supplementary Eq.~\ref{Eq:HDD} and Supplementary Eq.~\ref{Eq:HZNRot}. We typically use 16 different initial states, spanning the whole range of the total spin $z$ projections from $-N/2$ to $+N/2$. The same initial states are used through when sweeping the parameters of the Rf pulse sequence. Once the spin evolution is calculated for a particular Rf pulse sequence, we use the final wavefunction $\psi_{\rm{Fin}}$ to calculate the final nuclear spin polarization $I_{{\rm{z,Fin}}}=\langle \psi_{\rm{Fin}} \vert \sum_{k=1}^N \hat{I}_{{\rm{z}},k}\vert \psi_{\rm{Fin}}\rangle$. Each final polarization value is normalized by the sign of the initial polarization $I_{{\rm{z,Init}}}$, and these normalized values are then averaged over all the initial states $\psi_{\rm{Init}}$. Such averaging over multiple initial states helps to eliminate the oscillations that arise due to the small number of spins $N$ in the model. This averaging also mimics the experimental conditions, where the initial optically-induced nuclear polarization is subject to fluctuations due to fluctuations and drifts of various experimental parameters (e.g. optical pump power).

When modeling the nuclear spin dynamics of the homogeneous $I_{\rm{z}}=\pm1/2$ subspace we set all resonance offsets to be zero $\Delta \nu_k=0$. For the $I_{\rm{z}}=(-3/2,-1/2)$ satellite transition subspace we model the resonance offset of each nuclear spin $k$ as $\Delta \nu_k=\Delta \nu_0 (k-1) +\mathrm{N}(0,0.2\Delta \nu_0/(2\sqrt{2 \ln{2}}))$, where $\mathrm{N}(\mu,\sigma)$ stands for a normal distribution with the mean $\mu$ and standard deviation $\sigma$. This model ensures that the resonance frequency of each nucleus is detuned from the nearest frequency of another nucleus by $\Delta \nu_0$, on average. Each resonance shift is randomly varied (by a $\approx20\%$ fraction of $\Delta \nu_0$) to avoid periodic recurrences and oscillations. Once the set of randomized $\Delta \nu_k$ is generated, it is fixed and used throughout the different modeling runs. We have used three sets with $\Delta \nu_0=0.8$~kHz, $\Delta \nu_0=1.2$~kHz, and $\Delta \nu_0=1.6$~kHz. The calculated final nuclear spin polarizations $I_{{\rm{z,Fin}}}$ are averaged over these three sets, once again to combat the oscillations that arise from the small number of spins $N$ in the model.

The numerical modeling closely follows the experiments: the total free evolution time $T_{\rm{FreeEvol}}$ and the number of cycles  $n_{\rm{Cycles}}$ of the dynamical decoupling sequence are varied, to calculate the final nuclear spin polarization $I_{{\rm{z,Fin}}}$, averaged over initial states and resonance offsets as described above. The dependencies on $T_{\rm{FreeEvol}}$ yield decay curves that are similar to the experimental curves shown in Fig.~2c of the main text. Alternatively, the same data can be plotted as a function of the total evolution time $T_{\rm{EvolTot}}$ and the decoupling sequence period $T_{\rm{Cycle}}$, such as shown in Figs.~3c, d of the main text. The numerically modeled values of the final nuclear polarization $I_{{\rm{z,Fin}}}$ shown in Figs.~3c, d of the main text are normalized by the initial nuclear polarization $I_{{\rm{z,Init}}}$ averaged over all the initial states. This way, the normalized values are limited to the $[-1,+1]$ range, with $+1$ corresponding to no decay of the nuclear spin polarization (i.e. complete preservation through dynamical decoupling). The values close to $0$ can be interpreted as complete decoherence, whereas the negative values of the normalized nuclear spin echo amplitude are a sign of recurrences in the dynamics of a small nuclear spin ensemble. The corresponding experimental data, shown in Figs.~3a, b of the main text, is also normalized, but using the amplitude of the simple Hahn echo decoupling measured at short $T_{\rm{FreeEvol}}$, which approximates the initial nuclear spin polarization with good accuracy.



\end{document}